\documentclass[a4paper,10pt]{article}
\usepackage[utf8]{inputenc}

\usepackage{amsmath}
\usepackage{textcomp}
\usepackage[linesnumbered,ruled,vlined]{algorithm2e}

\usepackage{graphicx}
\usepackage[usenames,dvipsnames]{color}
\usepackage{amssymb}
\usepackage{tabularx}
\usepackage{enumitem}
\usepackage[hang,flushmargin]{footmisc}
\usepackage{caption}
\usepackage{soul}

\DeclareMathOperator*{\argmin}{\arg\min}
\newcommand{\AAB}{AAB}
\newcommand{\CCB}{CCB}
\newtheorem{thm}{Theorem}[section]
\newtheorem{example}{Example}[section]
\newtheorem{lemma}{Lemma}[section]
\newtheorem{definition}{Definition}[section]

\newtheorem{corollary}[thm]{Corollary}
\newtheorem{proposition}{Proposition}[section]
\newcommand{\proofends}{\vspace{-0.1in}\begin{flushright}
\mbox{}\hfill{$\square$} \end{flushright}}
\begin{document}
\title{An Incentive Compatible Multi-Armed-Bandit Crowdsourcing Mechanism with Quality Assurance}
\author{Shweta Jain, Sujit Gujar, Satyanath Bhat, Onno Zoeter, Y. Narahari}
\maketitle

\begin{abstract}
Consider a requester who wishes to crowdsource a series of identical binary labeling tasks to a pool of workers so as to achieve an
assured accuracy for each task,
in a cost optimal way.  The workers are heterogeneous with unknown but
fixed qualities and their costs are private.
The problem is to select for each task an optimal subset of  workers  so that
the outcome obtained after aggregating the labels from the selected workers guarantees a target accuracy level. The problem is a challenging one even in a
non strategic setting since the accuracy of aggregated label depends on unknown qualities.
We develop a novel multi-armed bandit (MAB) mechanism for solving this
problem. First, we propose a framework,
{\em Assured Accuracy Bandit (AAB)}, which leads to a MAB algorithm,
{\em Constrained Confidence Bound for a Non Strategic setting (CCB-NS)}. We derive an upper bound on the number of time steps the algorithm chooses a sub-optimal set that depends on the target accuracy level and true qualities. A more challenging situation arises when the requester not only has to
learn the qualities of the workers but also elicit their true costs. We modify the CCB-NS algorithm to obtain an adaptive exploration separated algorithm  which we call { \em Constrained Confidence Bound for a Strategic setting (CCB-S)}. CCB-S algorithm produces an ex-post monotone allocation rule and thus can be transformed into an
ex-post incentive compatible and ex-post individually rational mechanism that learns the qualities of the workers and guarantees
a given target accuracy level in a cost optimal way. We also provide a lower bound on the number of times any algorithm should select a sub-optimal set and we see that the lower bound matches our upper bound upto a constant factor. We provide insights on the practical implementation of this framework through an illustrative example and we show the efficacy of our algorithms through simulations.
\end{abstract}
\section{Introduction}
\noindent Consider a company that provides financial advice to a collection of clients on whether to invest in a particular security or not. In order to provide such advice to each client, the company has a pool of financial consultants. Gathering the opinion of as many consultants as possible and aggregating their opinions (for example, using majority voting)  increases the probability of providing a high accuracy advice, however, it also entails increased costs.  The company has two conflicting business requirements, firstly to keep the costs low, and secondly, to provide an advice that meets a minimum threshold accuracy. The individual financial consultants  have unknown skill sets (qualities) and their costs are typically private information. Since the consultants are strategic, they might report higher costs for their services. On the other hand, to meet an accuracy threshold, the company needs to learn the qualities of the consultants. Assuming the qualities of the consultants do not change with time, their qualities can be learnt by giving them homogeneous tasks.

As an abstraction to such problems, we consider a series of homogeneous binary labeling tasks. There is a pool of agents and each agent has different expertise or quality which is fixed but unknown. Since the tasks are homogeneous, the quality of an agent does not change from one task to another. In addition, each agent incurs a cost to perform the task and the cost of an agent is his private information and thus it can be strategically misreported. As a design objective, it is required that the final outcome, obtained by aggregating the answers from the selected agents, achieves a certain target accuracy. The target accuracy level parameter provides a handle on the trade off between cost and accuracy.  A high value of the target accuracy level enhances the probability  of getting the right answer but at the same time may call for a larger number of agents to be commissioned leading to raise in costs. Therefore, one can choose a suitable target accuracy level as per task sensitivity at hand. 
In a nutshell, the goal   is to select a subset of the strategic agents with minimum cost to achieve a desired accuracy 
level of the aggregated answer for each task, at the same time giving the right incentives to the agents so that they report their costs truthfully.

In the absence of strategic play or when the costs are known, the setting reduces to a machine learning problem. Here, the requester has to select an appropriate set of workers so as to minimize the costs while learning the qualities. Though the requester can learn the qualities of the workers over a period by observing their performance on similar tasks, selecting low quality workers repeatedly may incur significant costs. Thus, the requester faces a dilemma of {\it exploration} (where he has to learn the qualities of the workers) versus  {\it exploitation} (where he has to choose the workers optimally based on learnt qualities). A natural solution to this problem can be be explored using techniques developed for the multi-armed bandit (MAB) problem \cite{LAI85,AUER00}. 
In fact many existing works considered learning qualities in non strategic crowdsourcing settings \cite{LONG12,RAYKAR10}.
However, an important challenge in our setting is the need to ensure the accuracy constraint which in turn depends on unknown qualities. Thus, there is a need to develop a new framework to address the accuracy constraint. 

An additional challenge arises when the costs of the workers are private and the strategic workers try to manipulate the learning algorithm by misreporting their costs so as to benefit themselves. In the strategic version of the above problem, we have the additional task to elicit the true costs using a suitable mechanism. 
If qualities were known, a natural way to ensure truthfulness is
to use a classic mechanism such as the VCG (Vickery-Clarke-Groves) mechanism which satisfies many desirable game theoretic properties. However, since the qualities are not known and need to be learnt, a VCG mechanism cannot be applied directly \cite{BABAIOFF09}. Thus, we need to solve the problem of learning qualities while eliciting the costs simultaneously.  In short, we need to meld the techniques from machine learning and game theory that would ensure honest behavior of the workers while the requester learns the qualities. Often such mechanisms are referred to as \emph{Multi-Armed Bandit Mechanisms or simply MAB mechanisms} \cite{BABAIOFF09,BABAIOFF10,DEVANUR09,SUJIT12}. 
The above MAB mechanisms, however, are not designed to achieve a target accuracy level but to only learn the qualities. The MAB mechanism proposed in this paper induces honest behavior, learns qualities and also achieves required target accuracy.

\subsection{Contributions}
\noindent The above discussion highlights the need to design a new approach to solve the problem of selecting a subset of the strategic workers to achieve a target accuracy level in a cost optimal way. 
This paper solves such a problem for the first time by modeling this problem in the multi-armed bandit mechanism framework.
We consider two versions of this
problem: (1) a \emph{non strategic version} where the costs are known and the qualities of the workers
have to be learnt and (2) a \emph{strategic version} where the costs are to be truthfully elicited as well. 
In particular, the following are our contributions in this paper.
\begin{itemize}
\item 

We propose a novel framework, \emph{Assured Accuracy Bandit} (\AAB) where we formulate an optimization problem. 
\item We provide a lower bound on the regret that any MAB algorithm in the AAB framework has to suffer (Theorem \ref{thm:lower_bound}). 
\end{itemize}
We consider two versions of AAB.\\
\textbf{\sc{Non Strategic Version}}
\begin{itemize}
\item In this setting, we design a novel algorithm, which we call, Non Strategic Constraint Confidence Bound (\CCB-NS).
\item Though the true qualities are not known, our algorithm makes sure that the accuracy constraint is satisfied with high probability (Theorem \ref{thm:constraint}). 

\item We provide  an upper bound on the number of times the algorithm selects a suboptimal worker set for a given problem 
that depends on the target accuracy level and the true qualities (Theorem \ref{thm:bad_rounds_ns}). The upper bound achieved by our algorithm matches the lower bound upto a constant factor. 
\end{itemize}
\textbf{\sc{Strategic Version}}
\begin{itemize}
\item In the strategic version of this problem where workers may not report their costs truthfully, we modify the \CCB-NS algorithm to an adaptive exploration separated algorithm, which we call Strategic Constrained Confidence Bound (\CCB-S) and  prove that the allocation rule provided by the \CCB-S is ex-post monotone (Theorem \ref{thm:ex-post_monotone}) in terms of the cost.  

\item Given this ex-post monotone allocation rule, we adopt the existing techniques \cite{BABAIOFF10} to design an ex-post truthful and ex-post individually rational mechanism (Corollary \ref{cor:ic_mechanism}).

\item For a particular optimization problem, we extend the \CCB-S algorithm to the non-exploration separated algorithm by exploiting the specific structure of the optimization problem. For this, we also show the efficacy of our algorithms and compare the algorithms with a variant of $\varepsilon_t-$greedy algorithm \cite{AUER00} through simulations.
\end{itemize}

To the best of our knowledge, this is the first mechanism that learns the qualities of strategic agents (in this case crowd workers) who have 
costs as private information where a certain target accuracy level is achieved for each task. In general, MAB mechanisms are popular in the context of sponsored search auctions which are forward auctions. We extend the work to a crowdsourcing context  which is a reverse auction setting.

\subsubsection{Organization} The paper is organized as follows. We present a summary of the relevant work in Section \ref{sec:related_work}. In  Section \ref{sec:formulation}, we provide a general formulation of the problem. Next, we present our model in two different stages. First, we discuss the non strategic model in Section \ref{sec:non-strategic} and next in Section \ref{sec:mechanism_design}, we discuss the strategic version using  mechanism design. In Section \ref{sec:practical_aspects}, we provide an extension of the strategic version to a more practical setting where we provide conditions when an approximate solution of the optimization problem can be incorporated and workers can be eliminated in the strategic setting, thus avoiding higher cost in exploration steps. In this Section we also compare our algorithm with a variant of traditional MAB algorithm, $\varepsilon_t-$greedy algorithm through simulations.
Future work and conclusions are provided in Section \ref{sec:future_work}.

\section{Related Work}\label{sec:related_work}
First, we describe the state of the art addressing the non strategic versions of problems in crowdsourcing such as learning the qualities of the workers to improve the accuracy of the predicted answer. We then look into mechanism design literature in crowdsourcing. Our setting involves both learning and mechanism design. MAB mechanisms provide a natural solution in such setting. We also review relevant MAB problems and MAB mechanism design literature.
\subsubsection*{Learning in Crowdsourcing}
\noindent Ho et. al. \cite{HO13} considered a similar setting where an assured quality needs to be satisfied for each task. However, they dealt with a specific error probability function with a uniform and known cost of the workers. We address the heterogeneous setting with costs being privately held by strategic workers and we work with any general error probability function. Abraham et. al. \cite{ABRAHAM13} consider a setting where a certain accuracy is required to be met for a given micro-task. The authors considered the problem of aggregating answers in a sequential way until a certain accuracy is achieved.  Homogeneous workers are assumed in a cluster and thus the goal is to select a single optimal crowd for a single task. In a general setting, their assumption of a crowd having sufficient number of homogeneous quality workers may not hold. Our setting is more general where an optimal subset of workers (arms), with heterogeneous qualities, needs to be selected  at one go for a given micro-task. \cite{FAN15} consider a model where, workers have different quality for each variety of task. In order to assign a task to the worker, his quality on the previous similar tasks need to be estimated. However, an assured accuracy model is not considered.  Improving the quality of answers while minimizing the cost is considered by Karger et al. \cite{KARGER11} where the final answer is predicted using a low rank approximation method. Work by Raykar et al. \cite{RAYKAR10} considers learning a classifier while learning the qualities of the workers using EM \cite{DAWID79} algorithm. Viappiani et al. \cite{VIAPPIANI11} consider a Bayesian approach to learn the class label which take noisy observations from experts. Though, the models proposed \cite{VIAPPIANI11,DAWID79,RAYKAR10} work well experimentally, there are no analytic guarantees on the predicted outcome. Tran-Thanh et al. \cite{LONG12} present an MAB algorithm for efficient selection of  capacitated workers where each worker can perform only limited number of tasks. The authors formulated this as a knapsack problem. For each task, a single non strategic worker is selected as opposed to the subset selection of strategic workers, whose costs need to be elicited.

Though the literature addresses how to learn the quality of workers, none of the above papers addresses the challenge in meeting the target accuracy level on each task in a heterogeneous cost model. We also consider the strategic version where the costs can be misreported by the workers.

\subsubsection*{Mechanism Design in Crowdsourcing}

A majority of the literature on mechanism design in crowdsourcing involves design of pricing strategies with online workers. Babaioff et al. \cite{BABAIOFF12} use an MAB mechanism to determine an optimal pricing mechanism for a crowdsourcing problem having homogeneous qualities within a  specified budget (known as bandits with knapsack). Work by Singla and Krause \cite{SINGLA13} assumes costs to be private information and proposes a posted price mechanism to elicit the true costs from the users using MAB mechanisms while maintaining a budget constraint. Mechanism design in online procurement auctions  \cite{BABAIOFF12,SINGLA13,BADANADIYURU12,SINGER13} considers homogeneous quality workers. Our setting is more general where an auction mechanism is considered to elicit the true costs from the workers with heterogeneous qualities.

Garg et al. \cite{GARG12} and Bhat et al. \cite{SATYANATH14} consider the costs of the workers to be public and the qualities to be private and strategic. Another line of work involves incentivizing people to work with their true qualities, when the qualities are privately held by the workers \cite{WITKOWSKI13} in peer prediction markets. Cavallo and Jain \cite{RUGGIERO13} analyze crowdsourcing tasks as winner take it all auctions in game theoretic settings. They assume that only one worker gets paid and do not try to learn the qualities over period. \cite{GUJAR15a,GUJAR15b} adopt techniques from online mechanism design for eliciting the worker preferences but do not address the task accuracy problem.

Mechanism design theory typically has been used in crowdsourcing either to elicit the costs of the workers where qualities are homogeneous and known or to elicit the qualities of the workers assuming the costs to be known. Our work addresses the setting where the qualities of heterogeneous workers are to be learnt and the heterogeneous costs are to be elicited. 

\subsubsection*{MAB Algorithms}
A rich body of literature is available on the MAB problem. Our problem belongs to the  stochastic MAB setting, where the reward of each arm is fixed but unknown. A recent survey by  Bubeck and  Cesa-Bianchi \cite{BUBECK12} compiles several variations on stochastic and non-stochastic MAB problems. The setting that is closest to ours is considered by Shipra Agrawal and Nikhil Devanur \cite{SHIPRA14} where a general bandit problem with concave rewards and convex constraints is solved. Our problem setting is a further generalization, as the constraints in AAB are not convex. Moreover, the constraint is satisfied in expectation in \cite{SHIPRA14} as opposed to our work, where the constraint needs to be satisfied at each round. The Probably Approximately Correct (PAC) learning framework is considered in \cite{DAR06,SHIVARAM10,ZHOU14}. Our learning algorithm may appear closely related to the PAC learning setting but it differs in a subtle but important way. The solution obtained from any PAC algorithm is approximately correct with high probability after arms are pulled for a certain number of rounds, which depends on the provided approximation factor and the confidence. In our setting, the goal is to select an optimal set with high probability since a constraint needs to be satisfied with respect to stochastic qualities. Moreover, the number of exploration steps are adaptive that depends on the true qualities and the target accuracy level as opposed to the fixed number of exploration rounds in the PAC setting. The combinatorial MAB problem introduced by Chen, Wang, and Yuan \cite{CHEN13} is relevant to our work. Pure exploration strategy in combinatorial framework is considered in \cite{CHEN14} where the objective is to identify an optimal subset from given feasible subsets. However, in our setting collection of feasible sets is not given and has to be learnt over the time and this makes our work different from \cite{CHEN13,CHEN14} in the non strategic setting. A constrained MAB problem for single pull is discussed by Ding et. al. \cite{DING13}, where each arm is associated with random rewards and the goal of the algorithm is to maximize the reward such that total cost which is also stochastic in all rounds 
does not exceed the budget but the constraint is on overall rounds instead of each round.

\subsubsection*{MAB Mechanisms}

Multi-armed Bandit mechanisms in the forward setting, in particular, as applied to sponsored search auction are recent advancements that combine the area of MAB problems and mechanism design. Any deterministic truthful MAB mechanism must be exploration separated i.e. allocation in the learning phase should not depend on the bids and thus the regret of any such algorithm is at least $O(T^{2/3})$ where $T$ is the total number of rounds \cite{BABAIOFF09,DEVANUR09}. The results are also extended to multiple pull multi-armed bandits i.e. to the case of multiple slot sponsored search auction \cite{GATTI12,SUJIT12}. The techniques developed in these papers cannot be adopted to this setting because 1) the workers need to be paid in spite of their failure as opposed to the setting where a payment is made only if there is a success or a click 2) the setting is the constrained multi-armed bandit setting as opposed to the setting of traditional multi-armed setting where a best subset of arms is selected 
without constraints. Also note that, we are considering a reverse auction setting as opposed to the forward auction in the existing literature on MAB mechanisms. Babaioff et al. \cite{BABAIOFF10} design a general procedure which takes any monotone allocation rule as input and converts it into a randomized truthful mechanism which implements the input allocation rule with high probability and requires evaluation of the input allocation rule exactly once. As an application of this transformation, an MAB mechanism that is ex-post incentive compatible and ex-post individual rational with regret of $O(T^{1/2})$ is proposed. In our current work, we use this transformation and propose an ex-post monotone allocation rule in the case of a reverse auction in a constrained multi-armed bandit setting.
Mechanism proposed in \cite{SATYANATH15} can be translated in this setting to balance the trade-off between quality and the cost. However, the authors do not cater to the final accuracy of the task and selecting a single worker was considered as oppose to our model.

Our preliminary results appeared in \cite{SHWETA14} where we considered only a certain type of error probability function to ensure the target accuracy level. This current paper represent a significant improvement over our previous paper and the techniques developed in this paper are applicable to a general class of error probability functions satisfying monotonicity and bounded smoothness properties which we define later.

\section{The Model} 
\label{sec:formulation}
\begin{figure}[h!]
\centering
 \includegraphics[width=4.2in]{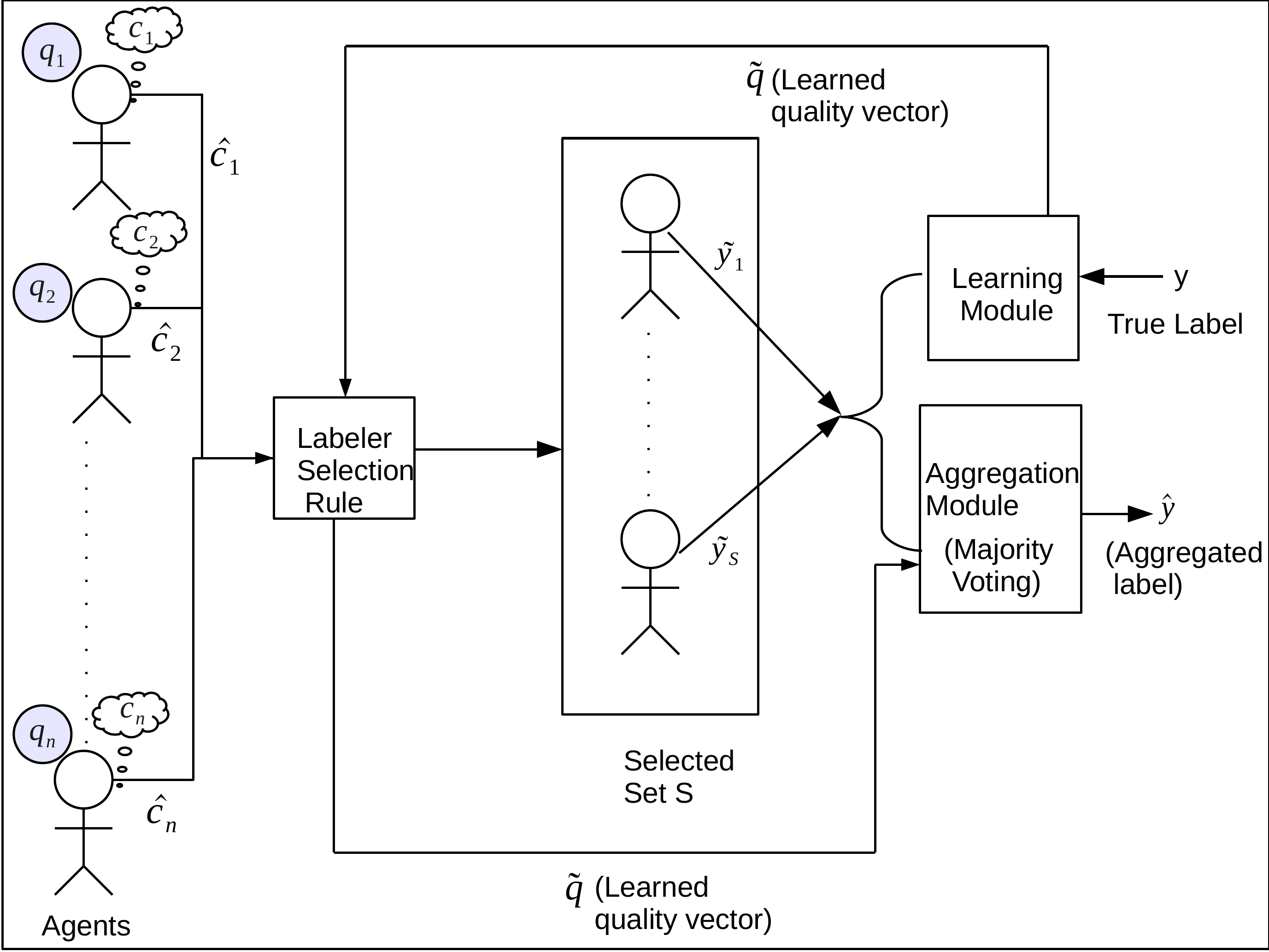}
\caption{The model: $q_i$, $c_i$ represent quality and true cost respectively of a worker $i$, $\hat{c}_i$ represents the reported cost by the worker $i$. The noisy label obtained from the worker $i$ is denoted by $\tilde{y}_i$.}
\label{fig:model}
\end{figure}

\noindent 
Let $\mathcal{N}$ be a set of $n$ crowdsourcing workers available
for working on $T$ homogeneous crowdsourcing tasks. Each agent or worker $i$ has an associated quality $q_i \in [0.5, 1]$, which represents the probability that the answer given by him is correct. By homogeneous or similar binary labeling tasks, we mean that each worker's quality is the same for all the tasks. We assume that the workers are not spammers and their quality of service is at least $0.5$. The quality of any worker $i$ is assumed to be independent of the qualities of other workers. A worker $i$ incurs a cost $c_i \in \mathbb{R}$ which is privately held and can be reported strategically by the workers. Let $1-\alpha$ be the target accuracy level ($\alpha$ is the threshold level) provided by the requester that determines the trade-off between the cost and the accuracy to be achieved for a particular task. We consider binary classification tasks where the labels are either zero or one. Our model is summarized in Figure~\ref{fig:model}.

\begin{table}[h!]
\caption{Notation Table}
\label{not_table}
\centering
\begin{tabular}{|l | l|}
\hline\hline
Notation & Description \\ [0.5ex]
\hline\hline
$\mathcal{N}$ & Set of workers available\\
$n$ & Number of workers available \\
$S$ & Set of workers selected\\
$S^t$ & Subset of workers selected for task $t$ \\
$T$ & Number of tasks\\
$t$ & Index for a task $t \in \{1,2,\ldots,T\}$ (also referred as rounds)\\
$\alpha$ & $(1-\alpha)$ is the target accuracy level required \\
$\mu$ & Confidence level required to satisfy accuracy constraint\\
$y^t$ & True label for task $t$\\
$\tilde{y}_i^t$ & Label reported by worker $i$ for task $t$\\
$\tilde{y}(S^t)$ & Vector of labels $\tilde{y}_i^t\ \forall i \in S^t$\\
$\hat{y}^t$ & Label predicted for task $t$ by aggregating all labels $\tilde{y}_i^t\ \forall i \in S^t$\\
$n_{i}(t)$ & Number of times worker $i$ is selected for tasks $1,2,\ldots,t$\\
$k_{i}(t)$ & Number of times worker $i$ got the label right for tasks $1,2,\ldots,t$\\
$q_i$ & True  quality of worker $i$ \\
$q$ & Quality vector $(q_1,q_2,\ldots,q_n)$\\
$\hat{q}_i(t)$ & Estimated quality of worker $i$ until tasks $t$\\
$\hat{q}_i^+(t)$ & Upper confidence bound (UCB) on $q_i$ until tasks $t$\\
$\hat{q}_i^-(t)$ & Lower confidence bound (LCB) on $q_i$ until tasks $t$\\
$c_i$ & True cost of worker $i$ for executing each task \\
$\hat{c}_i$ & Reported cost of worker $i$ \\
$c$ & Cost vector $(c_1,c_2,\ldots,c_n)$\\
$\hat{c}$ & Reported cost vector $(\hat{c}_1,\hat{c}_2,\ldots,\hat{c}_n)$\\
$C(S)$ & $\displaystyle\sum_{i \in S} c_i$\\
$S^*$ & An optimal set of workers with respect to true qualities \\
$f_S(q)$ & Error probability function of $S$ with $q$\\
$\Delta$ & Separation parameter of $f_S(q)$ from $\alpha$ for all set i.e. $|\alpha - f_{S}(q)| > \Delta\ \forall S \subseteq \mathcal{N}$ \\
$h(.)$ & Bounded smooth function s.t. $\displaystyle\max_i |q_i - q'_i| \le \delta \implies |f_S(q) - f_S(q')| \le h(\delta)\ \forall S \subseteq \mathcal{N}$\\
$L$ & Loss incurred when constraint is not satisfied\\
$R$ & Reward given when constraint is satisfied\\
\hline\hline
\end{tabular}
\end{table}

Notations are summarized in Table \ref{not_table}. The error on a task with inputs from the workers depends on the qualities of the workers and the rule to aggregate these answers. We abstract this as error probability function which we describe in the following subsection.
\subsection{Error Probability Function}
Let $f_S(q)$ be any function that represents the error probability (hence $(1 - f_S(q))$ captures the accuracy) when a set $S$ is selected with quality profile $q = (q_1,q_2,\ldots,q_n)$. The problem we seek to solve in this paper involves minimizing the cost, at the same time, satisfying the constraint that $f_{S}(q) < \alpha$ where $(1-\alpha)$ is the target accuracy level. Depending on the aggregation rule and the requester requirements, different error probability functions could be defined. Our framework and the solution approach are general and work for any error probability function that satisfies the following properties:

\begin{itemize}
\item {\em Monotonicity:} The error probability function $f_S(q)$ is said to be monotone if for all quality profiles $q$ and $q'$ such that if $\forall i \in \mathcal{N},\ q'_i \le q_i,\ \text{we have,}$
$$f_S(q') < \alpha \implies f_S(q) < \alpha,\ \forall S \subseteq \mathcal{N},\ \forall \alpha \in [0,1] \; .$$
That is, an increase in quality of each worker can only increase the accuracy or decrease the error probability.
\item {\em Bounded smoothness:} The error probability function $f_S(q)$ satisfies bounded smoothness if there exists a strictly increasing, continuous (hence, invertible) function $h$ such that if
$$\displaystyle\max_i |q_i - q'_i| \le \delta \implies |f_S(q) - f_S(q')| \le h(\delta),\ \forall S \subseteq \mathcal{N}, \forall q,q' \in [0.5,1]\; .$$
That is, when the two quality profiles are close, the difference in error probability function with respect to these quality profiles is bounded by a monotone continuous function $h$.
\end{itemize}

These properties are similar to the properties satisfied by the reward function in \cite{CHEN13} and are satisfied by various error probability functions. Next, we give certain examples of error probability functions that satisfy the properties of monotonicity and bounded smoothness when majority voting is used as an aggregation rule. Note that, the algorithm is general enough to incorporate any aggregation rule and any error probability function if monotonicity and bounded smoothness properties are satisfied.

\subsubsection{Examples of  error probability functions}

\noindent Let $S$ be the selected set with players $\{1,2,\ldots,s\}$ with the quality profile $q$, such that $q_1 \le q_2 \le \ldots \le q_s$ to whom we assign a certain task $t$. For notation convenience, we drop $t$ and let $\tilde{y}_i \in \{-1,1\}$ be the reported label that we get from the worker $i \in \{1,2,\ldots,s\}$ and $\tilde{y}(S) = (\tilde{y}_1, \tilde{y}_2,\ldots,\tilde{y}_s)$ be the vector of reported labels from the workers set $S$. Then, the predicted label $\hat{y}$ when a majority voting rule is used as an aggregation rule is given by:

\label{majority_voting}
\begin{align}
\label{eq_majority_voting}
\hat{y} =
\begin{cases}
1\ \text{if}  \displaystyle \sum_{i \in S} \tilde{y}_i > 0,\\
0\ \text{otherwise.}
\end{cases}
\end{align}

\begin{example}
The probability of the most likely outcome that leads to an error is given by \cite{SHWETA14}:
\begin{equation*}
\begin{aligned} 
&\mathbb{P}(E_{S(q)}) = \displaystyle \max_{\tilde{y}(S) \in \{-1,1\}^S} \mathbb{P}(\tilde{y}(S), \hat{y} \ne y|y)\\
&= \displaystyle \max_{\tilde{y}(S) \in \{-1,1\}^S}(\mathbb{P}(\hat{y} \ne y|y,\tilde{y}(S))\mathbb{P}(\tilde{y}(S)|y))\\
&= (1-q_1)(1-q_2)\ldots(1-q_{s'})q_{s'+1}\ldots q_s,\\
&\text{where} \ s' = \lfloor((s+1)/2) \rfloor.  
 \end{aligned}
\end{equation*}
Note that once $y,\tilde{y}(S)$ is fixed, $\hat{y} \ne y$ is either true or false and hence  
 $\mathbb{P}(\hat{y} \ne y|y,\tilde{y}(S))$=0 or 1. $\mathbb{P}(\tilde{y}(S)|y))$ is maximum when the top half quality workers make mistakes in which case, $\hat{y} \ne y$. 
Instead of satisfying the constraints with respect to $\mathbb{P}(E_{S(q)})$, one can satisfy the constraint with respect to the quantity $\hat{\mathbb{P}}(E_{S(q)})$ which is given as:
\begin{align}
f_S(q) = \hat{\mathbb{P}}(E_{S(q)}) = (1-q_1)(1-q_2)\ldots(1-q_{s'}) \label{monotone_worst_case_error}\; .
\end{align}
Note that $f_S(q)$ satisfies monotonicity as well as the  bounded smoothness property.
\end{example}
\begin{example}
\label{ex:min_knapsack}
The average probability of error is given by \cite{LI13}:
\begin{align*}
 \mathbb{P}(E_{S(q)}) &= \mathbb{P}(y=1)\mathbb{P}(\hat{y} = -1 | y = 1) + \mathbb{P}(y=-1)\mathbb{P}(\hat{y}=1|y=-1)\\
 &= \pi \mathbb{P}\left(\displaystyle\sum_{i=1}^s\tilde{y}_i \le 0 | y=1\right) + (1-\pi)\mathbb{P}\left(\displaystyle\sum_{i=1}^s\tilde{y}_i > 0|y=-1\right),
\end{align*}
where $\pi$ is the probability that true label $y$ is $1$.
Let us now focus on $\mathbb{P}\left(\displaystyle\sum_{i=1}^s \tilde{y}_i \le 0 | y=1\right)$.\\
\begin{align*}
 \mathbb{E}[\tilde{y}_i | y=1] = -\mathbb{P}(\tilde{y}_i=-1 | y=1) + \mathbb{P}(\tilde{y}_i=1 | y=1) = (2q_i - 1)\; .
\end{align*}
Now,
\begin{align*}
\mathbb{P}\left(\displaystyle\sum_{i=1}^s \tilde{y}_i \le 0 | y=1\right) &= \mathbb{P}\left(\displaystyle\sum_{i=1}^s \tilde{y}_i - \mathbb{E}\left[\sum_{i=1}^s\tilde{y}_i|y=1\right] \le - \mathbb{E}\left[\sum_{i=1}^s\tilde{y}_i|y=1\right] | y=1\right)\\
&=\mathbb{P}\left(\displaystyle\sum_{i=1}^s \tilde{y}_i - \mathbb{E}\left[\sum_{i=1}^s\tilde{y}_i\right] \le - \sum_{i=1}^s(2q_i - 1)\right)\\
&\le \displaystyle\exp\left(\frac{-\left(\displaystyle\sum_{i=1}^s(2q_i-1)\right)^2}{2\sum_{i=1}^s 1}\right)\ \text{(By Hoeffding's inequality)}.
\end{align*}
Similarly, it can be shown that:
$$\mathbb{P}\left(\displaystyle\sum_{i=1}^s \tilde{y}_i > 0 | y=-1\right) \le \displaystyle\exp\left(\frac{-\left(\displaystyle\sum_{i=1}^s(2q_i-1)\right)^2}{2\sum_{i=1}^s 1}\right)\; .$$
Assuming $\pi = 1-\pi = 0.5$, we get,
$$\mathbb{P}(E_{S(q)}) \le f_S(q) = \displaystyle\exp\left(\frac{-\left(\displaystyle\sum_{i=1}^s(2q_i-1)\right)^2}{2\sum_{i=1}^s 1}\right)\; .$$
Again, one can verify that function $f_S(q)$ is monotone and satisfies bounded smoothness property. If the quality of every worker $i$ satisfies $\frac{1}{2} + \epsilon \le q_i \le 1$, then, $(2q_i - 1) \ge 2\epsilon$ and the above expression can be simplified as:
$$\mathbb{P}(E_{S(q)}) \le \displaystyle\exp\left(\frac{-\left(\displaystyle\sum_{i=1}^s(2q_i-1)\right)^2}{2\sum_{i=1}^s 1}\right) \le \displaystyle\exp\left(-\displaystyle\sum_{i=1}^s(2q_i-1)\epsilon\right) = f_S(q).$$
\end{example}


From the above examples we see that it is reasonable to assume monotonicity and bounded smoothness for $f_S(q)$.

Now, we describe our framework in which the optimization problem takes the center stage.

\subsection{Assured Accuracy Bandit (\AAB) Framework}
Recall that a task $t \in \{1,\ldots,T\}$ needs to be completed with an assured accuracy provided by the requester with the optimal cost in a sequential fashion. Thus, for each task $t$, the goal of the requester is to select a set of workers $S^t$, such that the error probability function is less then the threshold level. At the same time, the requester has to make sure that the tasks are completed optimally in terms of costs. Hence for each task $t$, we need to solve the following
optimization problem.
\begin{equation}
\label{opt_problem}
\begin{array}{|c|}
\hline\\
\displaystyle \min_{X_i^t \in \{0,1\} } \displaystyle \sum_{i} c_iX_i^t,\\ 
\text{s.t.}, \\  
f_{\{i: X_i^t = 1\}}(q) < \alpha.\\
\;\\
\hline
\end{array}
\end{equation}
where the qualities of the workers are not known a priori and hence need to be learnt by giving tasks repeatedly to the workers. Also, solving the optimization problem, the requester has to make sure that the constraint in (\ref{opt_problem}) is satisfied with respect to the true qualities with high confidence. We refer to this novel framework as Assured Accuracy Bandits (AAB).

\subsubsection{Regret in AAB Framework}

\noindent The performance of any MAB algorithm is typically measured by the regret it achieves.
{\em Regret} in an MAB framework is defined to be the reward difference between the learning algorithm and the optimal algorithm. We will see later that our algorithm is designed in such a way that for each task $t$, the constraint given by (\ref{opt_problem}) is satisfied with probability $(1-\mu)$, where $\mu$ is the confidence parameter with which constraint is satisfied.
Thus we can define the regret of an algorithm $\mathcal{A}$ if the constraint is satisfied as:
{\allowdisplaybreaks
\begin{align} \label{regret_def}
& \mathcal{R}(\mathcal{A}) = \displaystyle \sum_{t=1}^T \displaystyle \sum_{i \in S^t} c_i - T\displaystyle\sum_{i \in S^*} c_i,\\
&S^t\ \text{is the set selected by the algorithm A for task}\ t, \nonumber \\
&S^* \ \text{is the optimal set with known qualities}. \nonumber
\end{align}
}
Since the constraint for each task is satisfied with probability $(1-\mu)$, we can also bound the total expected regret by:
\begin{align} \label{eq:regret_penalty}
\mathbb{E}[\mathcal{R}(\mathcal{A})] = (1-\mu)\displaystyle\left(\displaystyle \sum_{t=1}^T \displaystyle \sum_{i \in S^t} c_i - T\displaystyle\sum_{i \in S^*} c_i \right) + \mu L T\; ,
\end{align}
where $L$ is the cost that is incurred by the requester if the constraint fails to satisfy. We are considering a setting where value of $L$ is large, and the requester would not want to violate the constraint. However, due to stochasticity involved in learning the qualities, there is a small probability ($\mu$) with which the constraint can be violated.
With $\mu = 1/T$, we get the regret expression as:
\[\mathbb{E}[\mathcal{R}(\mathcal{A})] = (1-\frac{1}{T})\displaystyle\left(\displaystyle \sum_{t=1}^T \displaystyle \sum_{i \in S^t} c_i - T\displaystyle\sum_{i \in S^*} c_i \right) + L \;.\]

\subsection{Lower Bound on the Regret}
We first start with an important property called as $\Delta-$separated property that we assume any quality profile $q$ satisfies. The property is given as follows: 
\begin{definition}[$\Delta$-Separated Property:]
We say that $q$ is \textit{$\Delta$-Separated} with respect to the threshold $\alpha$ if $\exists \Delta>0$ such that, $\Delta = \inf_{S \subseteq \mathcal{N}} \ |f_{S}(q) - \alpha| $. That is, no set of workers, $S$, has probability of error $f_S(q) \in [\alpha-\Delta, \alpha+\Delta]$.
\end{definition}

Given a quality profile $q$ that satisfies $\Delta-$separated property, we now provide a lower bound on the regret that any algorithm in AAB framework has to suffer. 


\begin{thm}
 \label{thm:lower_bound}
Let $n_S(\mathcal{A})$ denotes the number of times a worker set $S$ is selected till time $T$ by any algorithm $\mathcal{A}$. Consider any algorithm that solves the optimization problem given by Equation (\ref{opt_problem}) and satisfies $E[n_S(\mathcal{A})] = o(T^a)\ \forall a > 0$ for any subset of worker $S$ which is not optimal. Then, the following holds:
$$\liminf_{T \rightarrow \infty} \mathbb{E}[\mathcal{R}(\mathcal{A})] \ge \frac{\ln T}{(h^{-1}(\Delta))^2},$$ where, $\Delta = \inf_{S \subseteq \mathcal{N}} \ |f_{S}(q) - \alpha|$ and $h(.)$ is the bounded smooth function.
\end{thm}
\textbf{Proof:}\\
The proof follows similar steps given in \cite{BUBECK12} for the lower bound proof of classical MAB problem for Bernoulli reward. For $p_1,p_2 \in [0,1]$, denote $kl(p_1,p_2)$ the Kullback-Leibler divergence between a Bernoulli of parameter $p_1$ and a Bernoulli of parameter $p_2$ defined as:

$$kl(p_1,p_2) = p_1 \ln \Big(\frac{p_1}{p_2}\Big) + (1-p_1)\ln\Big(\frac{1-p_1}{1-p_2}\Big).$$

It is easy to see that the function $x \mapsto kl(p_1,x)$ is a continuous function.

\begin{itemize}
 \item Consider two workers with quality profile, $q = (q_1,q_2)$ with $f_{\{1\}}(q) < \alpha < f_{\{2\}}(q)$ and $c_1 > c_2$. Since, $kl$ divergence is a continuous function and error probability function $f$ is monotone, for any $\epsilon > 0$, one can find quality profile $q' = (q_1, q_2')$ such that $f_{\{1\}}(q') < f_{\{2\}}(q') < \alpha$ and $kl(q_2,q_2') \le (1+\epsilon)kl(q_2, 1-\alpha)$. Thus, worker $1$ is optimal with quality profile $q$ but worker $2$ optimal with quality profile $q'$. Denote $\mathbb{P}, \mathbb{E}$ as the probability, expectation taken with respect to random variables generated with quality profile $q$ and $\mathbb{P}', \mathbb{E}'$ as the probability, expectation taken with respect to random variables generated with quality profile $q'$.
\item Denote $X_{2,1},X_{2,2}, \ldots, X_{2,T}$ as the sequence of successes obtained  when allocating tasks to worker $2$ where successes are coming from quality profile $q$. For any $t \in \{1,2,\ldots,T\}$, let
$$\hat{kl}_t = \sum_{i=1}^t \ln \frac{q_2 X_{2,i} + (1-q_2)(1-X_{2,i})}{q_2' X_{2,i} + (1-q_2')(1-X_{2,i})}.$$

Note that, $\mathbb{E}[\hat{kl}_{n_2(\mathcal{A})}] = n_2(\mathcal{A})kl(q_2,q_2')$. We also have the following change of measure identity for any event $B$ in the $\sigma-$algebra generated by $X_{2,1}, X_{2,2}, \ldots X_{2,T}$:
\begin{align}
\label{eq:measure}
  \mathbb{P}'(B) = \mathbb{E}[\mathbb{I}_B\exp(-\hat{kl}_{n_2(\mathcal{A})})].
\end{align}
\item Now, consider the event $C_T = \{n_2(\mathcal{A}) < \frac{1-\epsilon}{kl(q_2,q_2')}\ln(T)\ \text{and}\ \hat{kl}_{n_2(\mathcal{A})} \le (1-\frac{\epsilon}{2}) \ln(T)\}$. We will prove that $\mathbb{P}(C_T) \rightarrow 0$ as $T \rightarrow \infty$. From Equation (\ref{eq:measure}):
\begin{align*}
\mathbb{P}'(C_T) =  \mathbb{E}[\mathbb{I}_{C_T}exp(-\hat{kl}_{n_2(\mathcal{A})})] \ge e^{-(1-\epsilon/2)\ln(T)} \mathbb{P}(C_T).
\end{align*}
Let $f_T = \frac{1-\epsilon}{kl(q_2,q_2')}\ln(T)$. Then using Markov's inequality we have,

\begin{align*}
\mathbb{P}(C_T) \le T^{(1-\epsilon/2)} \mathbb{P}'(C_T) \le T^{(1-\epsilon/2)} \mathbb{P}'(n_2(\mathcal{A}) < f_T) \le T^{(1-\epsilon/2)} \frac{\mathbb{E}'(T - n_2(\mathcal{A}))}{T - f_T}.
\end{align*}

Since there are only two workers here, $T - n_2(\mathcal{A}) = n_1(\mathcal{A})$. With respect to quality profile $q'$ since worker $1$ is sub-optimal, we have:

\begin{align*}
 \mathbb{P}(C_T) \le T^{(1-\epsilon/2)} \frac{T^a}{T-f_T}, \ \forall a > 0.
\end{align*}

Consider $a < \epsilon/2$ then we get $\mathbb{P}(C_T)\rightarrow 0$ as $T \rightarrow \infty$.

\item Now, we will prove that $\mathbb{P}(n_2(\mathcal{A}) < f_T) \rightarrow 0$ as $T \rightarrow \infty$. We have,
\begin{align*}
\mathbb{P}(C_T) &\ge \mathbb{P}\left(n_2(\mathcal{A}) < f_T\ \text{and}\ \max_{t\le f_T}\hat{kl}_t \le \left(1-\frac{\epsilon}{2}\right)\ln(T)\right)\\
&= \mathbb{P}\left(n_2(\mathcal{A}) < f_T\ \text{and}\ \frac{kl(q_2,q_2')}{(1-\epsilon) \ln(T)}\max_{t\le f_T}\hat{kl}_t \le \frac{1-\epsilon/2}{1-\epsilon}kl(q_2,q_2')\right).
\end{align*}
Using the maximal version of the strong law of large numbers and since $kl(q_2,q_2') > 0$ and $\frac{1-\epsilon/2}{1-\epsilon} > 1$, we get:
\begin{align*}
 \lim_{T \rightarrow \infty} \mathbb{P}\left(\frac{kl(q_2,q_2')}{(1-\epsilon) \ln(T)}\max_{t\le f_T}\hat{kl}_t \le \frac{1-\epsilon/2}{(1-\epsilon)}kl(q_2,q_2')\right) = 1.
\end{align*}
Thus, from the previous point we get $\mathbb{P}(n_2(\mathcal{A}) < f_T) \rightarrow 0$ as $T \rightarrow \infty$. Thus, we get,

\begin{align*}
\mathbb{E}[n_2(\mathcal{A})] > \frac{1-\epsilon}{1+\epsilon}\frac{\ln(T)}{kl(q_2,1-\alpha)}. 
\end{align*}
Using the fact that $kl(p_1,p_2) \le \frac{(p_1-p_2)^2}{p_2(1-p_2)}$ and worker $2$ is suboptimal with quality profile $q$ we get:
\begin{align*}
\mathbb{E}[\mathcal{R}(\mathcal{A})] > \left(\frac{1-\epsilon}{1+\epsilon}\right)\frac{\ln(T)}{kl(q_2,1-\alpha)}(c_2 - c_1) \ge \left(\frac{1-\epsilon}{1+\epsilon}\right)\frac{\alpha(1-\alpha)\ln(T)}{(q_2 - (1-\alpha))^2}(c_2 - c_1).
\end{align*}

\item Since, $(1-\alpha)$ is the target accuracy, $f_S(1-\alpha) = \alpha$, for any subset $S$. From the bounded smoothness property, $|f_{\{2\}}(q) - f_{\{2\}}(1-\alpha)| \le h(q_2 - (1-\alpha)) \implies (q_2 - (1-\alpha))^2 \ge (h^{-1}(|f_{\{2\}}(q) - \alpha|))^2$. If we choose $q_1$ and $q_2$ such that $\Delta = |f_{\{2\}}(q) - \alpha|$ then we get $\mathbb{E}[\mathcal{R}](\mathcal{A})] > \left(\frac{1-\epsilon}{1+\epsilon}\right)\frac{\alpha(1-\alpha)\ln(T)}{(h^{-1}(\Delta))^2}(c_2-c_1)$, yielding the lower bound.
\end{itemize}
\proofends 

The above theorem proves that, in order to reach to the optimal solution in \AAB\ framework, it is required to pull a sub-optimal arm atleast $O\left(\frac{\ln(T)}{(h^{-1}(\Delta))^2}\right)$ number of times. Here, $\Delta$ depends on the problem instance.

\subsection{Related MAB Algorithms}

There are various MAB algorithms for different related settings. Here we list the two algorithms closest to our setting. 

\subsubsection{The UCB Algorithm}

The UCB algorithm proposed by Auer et al. \cite{AUER00} works for the classical MAB problem where only one arm is pulled at any given time. This algorithm maintains an upper confidence bound (hence the name UCB) on each arm which depends on its empirical reward as well as on the exploration factor to give the arm enough number of pulls. The algorithm achieves a regret of $O(\ln T)$ and it is the best possible regret that can be achieved. One extension to the multiple pull setting is to pull the arms in the increasing order of their upper confidence bound.

\subsubsection{The CUCB Algorithm}

The CUCB algorithm \cite{CHEN13} generalizes UCB to the combinatorial setting. At each time, a subset of arms is pulled and the rewards of all the selected arms are revealed. The algorithm works for general non-linear reward functions as long as monotonicity and bounded smoothness properties are satisfied by the reward functions. The general idea of the algorithm is to select the subset such that the reward is maximized with respect to the upper confidence bounds of all the arms similar to the UCB algorithm where a single arm is selected with the highest upper confidence bound. The CUCB algorithm achieves a regret of $O(\ln T)$. The \AAB\ framework handles an unknown stochastic constraint given by $f_S(q) < \alpha$ in contrast to the CUCB algorithm where the reward function is stochastic with known constraints.

We now provide some realistic assumptions that we have made in our model to design algorithms for \AAB\ framework.

\subsection{Assumptions}
\begin{itemize}
 \item We consider a series of binary classification tasks as an abstraction to our problem. 
 \item We assume that the error probability function satisfies the assumptions of monotonicity and bounded smoothness. These assumptions are natural and are satisfied by many interesting error probability functions.  
 \item We assume that the true label is observed once the task is completed. To motivate this assumption, we recall the trading example given in the introduction section where the company and the clients can realize the true label by the end of the day, for example, in an intraday trading.
 \item We assume that if all the workers are selected, then the constraint is always satisfied with respect to the true qualities. Thus, if qualities are only partially learnt, then the algorithm can select the complete set and satisfy the constraint. This is equivalent to saying that there are enough good workers.
\end{itemize}

\section{Non Strategic Version} \label{sec:non-strategic}
\noindent In this setting, our goal is to learn the qualities of the workers while assuming costs to be publicly known. We solve a general optimization problem given by Equation (\ref{opt_problem}) with unknown qualities which are to be learnt over a period of time. Since the workers obtain their labels according to the true qualities, the constraint has to be satisfied with respect to the true qualities. Since these qualities are unknown to the requester, he has to make sure that the constraint is satisfied with high probability. Note that our algorithm works in a general setting with any aggregation rule and with any error probability function that satisfies the monotonicity and the bounded smoothness properties. Thus, the algorithm uses the aggregation rule as a black box. 

\begin{definition}[Aggregate]
\label{def:aggregate}
An aggregate function takes the noisy labels of the selected set as input and produces a label $\hat{y}$ which best captures the opinion of labeler set. We call this label as the aggregated label. The aggregate function should ensure that the resulting error probability function satisfies the properties of monotonicity and bounded smoothness. For example, if the majority voting rule is used, the aggregated label $\hat{y}$ is computed using the equation (\ref{eq_majority_voting}).
\end{definition}

We now present the non strategic version of the Constrained Confidence Bound algorithm (CCB-NS) that satisfies the constraint for each task with high probability.

\subsection{\CCB-NS Algorithm} \label{sec:ccb_ns}
\begin{algorithm}[h!]
\DontPrintSemicolon
\caption{\CCB-NS Algorithm}
\label{alg:ccb_ns}
\KwIn {Set of workers $\mathcal{N}$, number of tasks $T$, parameter $\alpha$, confidence level $\mu$}

\KwOut {Labeler selection set $S^t$, Label $\hat{y}^t$ for all tasks $t \in \{1,2,\ldots,T\}$}
$\forall i \in \mathcal{N}$, $\hat{q}_i^+ = 1$, $\hat{q}_i^- = 0.5$, $k_{i}(1) = 0$ // Initialize UCB and LCB on qualities\;

$S^1 = \mathcal{N}$ // Select all workers initially \label{initialize} \;

Observe $\tilde{y}(S^1)$ and $\hat{y}^1 = \text{AGGREGATE}(\tilde{y}(S^1))$ (Definition \ref{def:aggregate})\; 

Observe true label $y^1$ \;

$\forall i \in \mathcal{N}$, $n_{i}(1) = 1$, $k_{i}(1) = 1$ if $\tilde{y}_i^1 = y^1$ and $\hat{q}_i = k_{i}(1)/n_{i}(1)$ \; 
$t = 2$ \;
$S^t = \displaystyle\argmin_{S \subseteq \mathcal{N}} \displaystyle\sum_{i \in S} c_i\ \text{s.t.}\ f_S(\hat{q}^+) < \alpha$ \;
\While{$f_{S^t}(\hat{q}^-) > \alpha$} 
{ \label{check_constraint}  // Explore (not the optimal set, add more workers to satisfy the constraint) \;
    $S^t = S^t \cup \text{MINIMAL}(S^t, \mathcal{N} \setminus S^t, \hat{q}^-)$  \label{additional_set} \;
    Observe labels of selected labelers $\tilde{y}(S^t)$ \;
    $\hat{y}^t = \text{AGGREGATE}(\tilde{y}(S^t))$\;
    Observe true label $y^t$ \; 
    \For{$i \in S^t$}
	{$n_{i}(t) = n_{i}(t-1) + 1$\; 
    \If{$\tilde{y}_i^t = y^t$}{$k_{i}(t) = k_{i}(t-1) + 1$}
    
    $\hat{q}_i = k_{i}(t)/n_{i}(t)$, $\hat{q}_i^+ = \hat{q}_i + 
    \sqrt{\frac{1}{2n_{i}(t)} \ln(\frac{2n}{\mu})}$, $\hat{q}_i^- = \hat{q}_i 
    - \sqrt{\frac{1}{2n_{i}(t)} \ln(\frac{2n}{\mu})}$ \label{update} \;}
    $t = t+1$\;
    $S^t = \displaystyle\argmin_{S \subseteq \mathcal{N}} \displaystyle\sum_{i \in S} c_i\ \text{s.t.}\ f_S(\hat{q}^+) < \alpha$ \label{selected_set} \;}
	$t^* = t$ \label{else_part} \;
        $S^{t^*} = S^t$ \label{return_opt_set}\;

\For{$t = t^* + 1$ to $T$\label{start_exploitation}} {
// Exploit (optimal set with high probability)\;
	$S^t = S^{t^*}$ \;
	Observe labels of selected labelers $\tilde{y}(S^t)$ \;
	$\hat{y}^t = \text{AGGREGATE}(\tilde{y}(S^t))$ \label{exploitation_step} 
}
\vspace{0.3cm}
Subroutine: MINIMAL$(S^t,S,q)$ \nonumber \;
Return a minimal set $S' \subseteq S$ of workers such that $f_{S^t \cup S'}(q) < \alpha$\;
If no such set $S'$ exists then return $S$
\end{algorithm}

\noindent The \CCB-NS algorithm (presented in Algorithm \ref{alg:ccb_ns}) works on the principle of the UCB algorithm \cite{AUER00} and ensures that the constraint in (\ref{opt_problem}) is satisfied with high confidence $\mu$. Input to the algorithm is parameter $\alpha$, the target accuracy (which is assumed to be same for all the tasks), the number of tasks $T$, the number of workers $n$, and confidence level $\mu$ with which the  constraint in (\ref{opt_problem}) is required to be satisfied. The output of the algorithm will be the subset $S^t$ and predicted label $\hat{y}^t$ for each task $t$. The predicted label $\hat{y}^t$ is decided based on an aggregation function (AGGREGATE) defined in \ref{def:aggregate} with noisy labels collected from the worker set $S^t$ as input.

Initially all the workers are selected to have some estimate about the qualities (Step \ref{initialize}). Their reported labels are aggregated and a label is predicted. Next, the algorithm observes the true label and updates the mean quality estimates, the upper and lower confidence bounds. Let $n_{i}(t)$ denote the number of times the $i^{th}$ worker is assigned the task and $k_{i}(t)$ denote the number of times the worker has provided the correct label up to the task $t$. Similar to the UCB algorithm \cite{AUER00}, the algorithm maintains upper confidence and lower confidence bounds on qualities. These bounds are given as follows:\\
 $$ \hat{q}_i^+(t) = \hat{q}_i(t) + \sqrt{\frac{1}{2n_{i}(t)} \ln\left(\frac{2n}{\mu}\right)},\ \hat{q}_i^-(t) = \hat{q}_i(t) - \sqrt{\frac{1}{2n_{i}(t)} \ln\left(\frac{2n}{\mu}\right)},\ \text{where}\ \hat{q}_i(t) = \frac{k_{i}(t)}{n_{i}(t)}\; .$$
 
 
 By Hoeffding's inequality, one can prove that the true quality $q_i$ lies between $\hat{q}_i^-(t)$ and $\hat{q}_i^+(t)$ with probability $1-\frac{\mu}{n}$ for any task $t$ and for any worker $i$. The bounds used by UCB1 algorithm given in \cite{AUER00} is given by:
 $$ \hat{q}_i^+(t) = \hat{q}_i(t) + \sqrt{\frac{2 \ln t}{n_{i}(t)}},\ \hat{q}_i^-(t) = \hat{q}_i(t) - \sqrt{\frac{2 \ln t}{n_{i}(t)}}.$$
 Since, we want to extend the algorithm to the strategic setting, we are using a constant term $\frac{2n}{\mu}$ in the bounds instead of $t$. 
 We initialize $\hat{q}_i^+(t)$ and $\hat{q}_i^-(t)$ by $1$ and $0.5$ respectively as the true qualities of the workers lie between $[0.5,1]$. In the algorithm, we represent $\hat{q}_i$, $\hat{q}_i^+$ and $\hat{q}_i^-$ to be the estimates till $t$ number of tasks.
The key idea in our algorithm is, till we have identified the optimal subset of workers, we solve the optimization problem using the upper confidence bound on the qualities which gives a cost effective subset. However, this subset need not meet the desired accuracy. Hence we add another subset of the workers from the remaining workers (using subroutine MINIMAL) that combined together ensures that the target accuracy is met even when we use the lower estimates, that is $\hat{q}^-$ in the constraints.
The fact that $q_i \geq \hat{q}_i^-$ with probability at least $1-\frac{\mu}{n}$ and the monotonicity of the error function ensures that the target accuracy level is achieved in each round with high probability.
Once the algorithm finds a subset that is optimal with respect to the upper confidence on qualities and achieves the target accuracy even when using the lower confidence on qualities, the algorithm stops learning and uses this set for the remaining tasks. We prove (Lemma \ref{opt_set}) that this is the required optimal set with high probability. Note that, in Step \ref{additional_set} if the MINIMAL function cannot find a set satisfying the target accuracy level using the lower confidence bound, then it simply returns $\mathcal{N}$ which meets the target accuracy level by our assumption.

We first see that the algorithm CCB-NS satisfies the constraint at each round with high probability. Note that by Hoeffding's inequality, for each $i$, $\hat{q}_i^- \le q_i \le \hat{q}_i^+$ with probability $1-\frac{\mu}{n}$. Since the workers make error independently, we have, $\forall i \in \mathcal{N} \ \hat{q}_i^- \le q_i \le \hat{q}_i^+$ with probability $\left(1 - \frac{\mu}{n}\right)^n \ge (1-\mu)$ by Bernoulli's inequality. Thus, $\forall i \in \mathcal{N} \ \hat{q}_i^- \le q_i \le \hat{q}_i^+$ with probability greater then $1 - \mu$. For brevity of notation, in the rest of the paper we will use $\hat{q}^- \le q \le \hat{q}^+$ to represent $\hat{q}_i^- \le q_i \le \hat{q}_i^+\ \forall i \in \mathcal{N}$.

From this, we make an important observation, 
as with probability at least, $1-\mu, \hat{q}^- \le q \le \hat{q}^+$ and monotonicity of
$f(.)$,
\begin{equation}
\mbox{w.p. at least }1-\mu, f_S(\hat{q}^+) < f_S(q) < f_S(\hat{q}^-)\ \forall S \subseteq \mathcal{N}
\label{eq:f_mono}
\end{equation}

\begin{thm}
\label{thm:constraint}
The CCB-NS algorithm satisfies the constraint in Equation (\ref{opt_problem}) with probability at least 1-$\mu$ at every round $t$.
\end{thm}
\noindent \textbf{Proof:}
\begin{itemize}
\item By our assumption that if all the workers are selected, the constraint is always satisfied, thus, in the rounds in which all the workers are selected, the constraint is satisfied.
\item Now, if set $S^t$ is returned by CCB-NS, then,\\
$f_{S^{t}}(\hat{q}^-) < \alpha \implies f_{S^{t}}(q) < \alpha$ with probability $1-\mu$ (From Equation \ref{eq:f_mono}). 
\end{itemize}
\proofends
We now show that if the algorithm exits the while loop in Step \ref{else_part} then the set $S^{t^*} = S^*$ (the optimal set) with probability at least $1-\mu$. For simplicity, in the rest of the paper, we assume that there exists a unique optimal set $S^*$, though the results can be easily generalized to the case where there are multiple optimal sets.
\begin{lemma}
 \label{opt_set}
Set $S^{t^*}$ returned by the \CCB-NS  algorithm is an optimal set with probability (w.p.) at least $1-\mu$. That is, $C(S^{t^*}) = C(S^*)$ w.p. $1-\mu$.
\end{lemma}
\noindent \textbf{Proof:}
Let $t^*$ be the round in which CCB-NS stops exploring. At $t=t^*$, 
\begin{itemize}
 \item Since $f_{S^*}(q) < \alpha$, we have, $f_{S^*}(\hat{q}^+) < \alpha$ with probability $1-\mu$ (From Equation (\ref{eq:f_mono})).
\item Let $S^{t^*}$ be the set of workers selected by CCB-NS.
As CCB-NS solves the optimization problem \ref{opt_problem}, 
$$C(S^{t^*}) \le C(S^*)\; .$$
\end{itemize}
At $t^*$,
\begin{align*}
f_{S^{t^*}}(\hat{q}^-) < \alpha &\Rightarrow f_{S^{t^*}}(q) \le f_{S^{t^*}}(\hat{q}^-) < \alpha\ \mbox{ with probability at least } 1-\mu,\\
&\Rightarrow C(S^{t^*}) = C(S^*)\; .
\end{align*}
\proofends

\subsection{Regret Analysis of \CCB-NS}

In this section, we aim to bound the number of non-optimal rounds for the \CCB-NS\ algorithm presented in Algorithm \ref{alg:ccb_ns}. 

\begin{definition}[Non-optimal Subset]
We say that at round $t$, a set $S^t$ selected by the algorithm is a \textit{non-optimal subset}, if $S^t \ne S^*$.
\end{definition}
\begin{definition}[Non-optimal Round:]
We say a round $t$ is a \textit{non-optimal round} if the selected set $S^t$ is not the optimal set $S^*$.
\end{definition}

We bound the number of exploration steps. Since the algorithm selects a set which satisfies the constraint for each task with high probability, we can bound the overall regret by bounding the number of rounds in which the algorithm selects a sub-optimal set $S^t$ i.e. $C(S^t) > C(S^*)$. If $C(S^t) = C(S^*)$, then we get zero regret for those rounds with probability $(1-\mu)$. We will show that the number of non-optimal rounds depends on the value of $\Delta$ where $\Delta = \inf_{S \subseteq \mathcal{N}} \ |f_{S}(q) - \alpha|$. The value of $\Delta$ is typically unknown to the requester since qualities are unknown but our algorithm does not require the value of $\Delta$ beforehand and thus, \CCB-NS is adaptive in nature. 

\begin{lemma}
\label{lemma1_non_strat}
If $\forall i \in \mathcal{N}$, number of times a worker $i$ is selected till tasks $t$, $n_{i}(t) \ge \frac{2}{(h^{-1}(\Delta))^2}\ln(\frac{2n}{\mu})$, then for any task $t$,
\begin{enumerate}
\item $\forall S \subseteq \mathcal{N},\ S\neq S^*$, $f_S(q) > \alpha \implies f_S(\hat{q}^+) > \alpha$ with probability $1-\mu$.
\item $f_{S^*}(\hat{q}^-) < \alpha$ with probability $1-\mu$.
\end{enumerate}
\end{lemma}
\noindent\textbf{Proof:}
Let $l = \frac{2}{(h^{-1}(\Delta))^2}\ln(\frac{2n}{\mu})$.
\begin{itemize}
\item By Hoeffding's inequality, $\hat{q}_i^+ - q_i \le 2\sqrt{\frac{1}{2n_{i}(t)}\ln(\frac{2n}{\mu})} \le 2\sqrt{\frac{1}{2l}\ln(\frac{2n}{\mu})}\ \forall n_i(t) \ge l,$ with probability $1-\frac{\mu}{n}$.
\item Substituting $l = \frac{2}{(h^{-1}(\Delta))^2}\ln(\frac{2n}{\mu})$, $\hat{q}_i^+ - q_i \le h^{-1}(\Delta)$ with probability $1-\frac{\mu}{n}$. Thus, $\hat{q}^+ - q \le h^{-1}(\Delta)$ with probability $1-\mu$.
\item By bounded smoothness and monotonicity, $f_S(q) - f_S(\hat{q}^+) \le h(h^{-1}(\Delta)) \le \Delta$ with probability $1-\mu$.
\item Similarly, $\forall S \subseteq \mathcal{N},\ f_S(\hat{q}^-) - f_S(q) \le \Delta$ with probability $1-\mu$.
\end{itemize}
Thus, $f_S(\hat{q}^+) \ge f_S(q) - \Delta$ and $f_{S^*}(\hat{q}^-) \le f_{S^*}(q) + \Delta$.\\
Thus, $f_S(q) > \alpha \Rightarrow f_S(\hat{q}^+) > \alpha\ (\because f_S(q) > \alpha + \Delta,  \Delta$-separated property).\\
And, $f_{S^*}(\hat{q}^-) < \alpha\ (\because f_{S^*}(q) < \alpha - \Delta,  \Delta$-separated property).
\proofends

\begin{lemma}
\label{lemma:non_strategic}
If a non-optimal set $S^t$ is selected for the task $t$ then there exists a worker $i \in S^t$ such that $n_{i}(t) \le \frac{2}{(h^{-1}(\Delta))^2}\ln(\frac{2n}{\mu})$ with probability $1-\mu$.
\end{lemma}
\textbf{Proof:} A non-optimal subset $S^t$ could be selected in two ways:
\begin{itemize}
\item $f_{S^t}(\hat{q}^+) < \alpha$ but $f_{S^t}(q) > \alpha$,
\item $f_{S^*}(\hat{q}^-) > \alpha$.
\end{itemize}
From Lemma \ref{lemma1_non_strat}, if $n_{i}(t) \ge \frac{2}{(h^{-1}(\Delta))^2}\ln(\frac{2n}{\mu}) \ \forall i \in S^t$, then, both the conditions are violated and thus a non-optimal subset is not selected. 
\proofends
\begin{thm}
\label{thm:bad_rounds_ns}
The number of non-optimal rounds by the \CCB-NS algorithm is bounded by $\frac{2n}{(h^{-1}(\Delta))^2}\ln(\frac{2n}{\mu})$ with probability $1-\mu$.
\end{thm}
\textbf{Proof:}
\begin{itemize}
\item Lemma \ref{opt_set} shows that the CCB-NS exploitation rounds are optimal rounds.
\item A new parameter $u_{i}(t)$ is associated with each worker. Whenever a set $S^t$ is selected then, $u_{i}(t) = u_{i}(t) + 1\ \text{s.t}\ i \in S^t\ \text{and}\ i = \displaystyle\argmin_{j \in S^t}u_{j}(t)$.
\item Every time a non-optimal subset $S^t$ is selected, $u_{i}(t)$ of only one worker is updated with the lowest value of $u_{i}(t)$ so far, such that $i \in S^t$. Thus, $u_{i}(t) \le n_{i}(t)\ \forall i \in \mathcal{N}\ \forall t \in \{1,\ldots,T\}$.
\item Thus, from Lemma \ref{lemma:non_strategic}, the number of exploration rounds is bounded by $\frac{2n}{(h^{-1}(\Delta))^2}\ln(\frac{2n}{\mu})$ with probability $1-\mu$.
\end{itemize}
Hence the theorem follows.
\proofends

\begin{corollary}
The total expected regret is bounded by
$$\left(1-\frac{1}{T}\right)\frac{2n}{(h^{-1}(\Delta))^2}\ln(2nT)C(\mathcal{N}) + L,$$
where $L$ is the loss incurred by the requester if the constraint is not satisfied.
\end{corollary}

The above corollary can be obtained by substituting $\mu = 1/T$. We see that the regret by CCB-NS algorithm matches the lower bound in AAB framework up to a constant factor.
In this section, we assumed that the costs are known. However, in real world situations, we need to elicit them truthfully from the strategic workers which we address next.

\section{Strategic Version} 
\label{sec:mechanism_design}

\noindent 
Without proper incentives, the strategic agents may not report their costs truthfully. In this section, we propose an algorithm which we call, CCB-S that satisfies a certain monotonicity property and then present a mechanism design implementation. We will first describe the game theoretic version of the problem addressed in the previous section. 

\subsection{The Model}
\noindent Denote the true cost of a worker $i$ by $c_i$ and the reported cost by $\hat{c}_i$. The valuation of a worker $i$ is given by $v_i = -c_i$. We denote the requester as agent $0$ and the valuation of the requester, when the task is allocated to the worker set $S$, by:
\[
    v_0(S)= 
\begin{cases}
    R & \text{if } f_S(q) < \alpha,\\
    -L              & \text{otherwise}.
\end{cases}
\]
Here, $f_S(q)$ is the error probability function that satisfies monotonicity and bounded smoothness properties. We denote $1-\alpha$ as the target accuracy level. The parameter $R$ denotes the reward that the requester gets for satisfying the constraint and $L$ denotes the loss he incurs if the constraint in Equation (\ref{opt_problem}) is not satisfied. Note that the requester is not considered to be strategic.
Social welfare $W(S)$ is given by:
\[
    W(S)= 
\begin{cases}
    R - \displaystyle \sum_{i \in S} c_i & \text{if } f_S(q) < \alpha,\\
    -L - \displaystyle \sum_{i \in S} c_i              & \text{otherwise}.
\end{cases}
\]

A mechanism $\mathcal{M}$ is denoted by the pair $(\mathcal{A},\mathcal{P})$, where $\mathcal{A} = (\mathcal{A}_1,\mathcal{A}_2,\ldots,\mathcal{A}_n)$ is the allocation vector where $\mathcal{A}_i$ represents number of tasks allocated to worker $i$ and $\mathcal{P} = (\mathcal{P}_1,\mathcal{P}_2,\ldots,\mathcal{P}_n)$ is the payment vector where $\mathcal{P}_i$ denotes the total payment made to the worker $i$  which depends on the reported cost profile $\hat{c}$. We work in a quasi-linear setting where the utility of every agent is given by:
\[u_i(c_i, \hat{c};q) = -c_i.\mathcal{A}_i(\hat{c};q) + \mathcal{P}_i(\hat{c};q)\; .\]

We consider the problem where a heavy penalty is incurred for providing the wrong answer and thus, the parameter $L$ is large.
We now review some of the desirable properties that the mechanism $\mathcal{M}$ should satisfy:

\begin{definition}[Incentive Compatible]
A mechanism $\mathcal{M}=(\mathcal{A},\mathcal{P})$ is said to be Incentive Compatible if reporting true valuations is a dominant strategy for all the workers. That is, $\forall i \in \mathcal{N}$,
\begin{align*}
&-c_i\mathcal{A}_i(c_i,\hat{c}_{-i};q) + \mathcal{P}_i(c_i,\hat{c}_{-i};q) \ge -c_i\mathcal{A}_i(\hat{c}_i, \hat{c}_{-i};q) + \mathcal{P}_i(\hat{c}_i, \hat{c}_{-i};q)\; \forall  \hat{c}_i \in [0,1] \mbox{ and } \hat{c}_{-i} \in [0,1]^{n-1} \; .
\end{align*}
\end{definition}

\begin{definition}[Individual Rationality]
 A mechanism $\mathcal{M}=(\mathcal{A},\mathcal{P})$ is said to be individually rational for a worker if participating in the mechanism always gives him positive utility. That is, $\forall i \in \mathcal{N}$,
 \begin{align*}
-\hat{c}_i\mathcal{A}_i(\hat{c}_i,c_{-i};q) + \mathcal{P}_i(\hat{c}_i,c_{-i};q) \ge 0\; \forall  \hat{c}_i \in [0,1] \mbox{ and }\ \forall c_{-i} \in [0,1]^{n-1}\; .
\end{align*}
\end{definition}

An important characterization for incentive compatible mechanisms provided by Myerson \cite{MYERSON81} states that for a mechanism to be truthful, the allocation rule should be monotone in terms of reported bids by the players. Babaioff et. al.
\cite{BABAIOFF10} provide a generic transformation that takes any monotone allocation rule and outputs a mechanism which is incentive compatible and individually rational. We can use this generic transformation to design the mechanism in our setting. We first provide the definition of monotonicity in our setting.

\begin{definition}[Monotonicity of Allocation Rule]
An allocation rule $\mathcal{A}$ is monotone if for every worker $i$, and for every fixed $\hat{c}_{-i} \in [0,1]^{n-1}$,
\begin{align*}
\hat{c}_{i} \le \hat{c}_{i}' \Rightarrow \mathcal{A}_i(\hat{c}_i, \hat{c}_{-i};q) \ge \mathcal{A}_i(\hat{c}_i', \hat{c}_{-i};q), \end{align*}
where $\mathcal{A}_i(\hat{c}_i,\hat{c}_{-i};q)$ is the number of tasks given to the $i^{th}$ worker with bids $\hat{c}_i$ and $\hat{c}_{-i}$.
\end{definition}

Since there is randomness involved due to learnt qualities, let us first define every possible random seed. The random variables are the labels provided by the workers and can affect the learnt qualities and thus the allocation rule.

\begin{definition}[Success Realization]
A success realization is a matrix $\rho \in \{0,1,-1\}^{n \times T}$ such that,
\[
\rho_{it}  = 
\begin{cases} 1 \; \mbox{if } \tilde{y}_i^t = y^t,\\
 0 \; \mbox{if } \tilde{y}_i^t \neq y^t, \\
 -1 \; \mbox{if worker }i\mbox{ is not selected for the }t^{th}\mbox{task}.
 \end{cases}
\]
\end{definition}

We also define relevant weaker notions of incentive compatibility, individual rationality and monotonicity. Note that, the allocation and payment rule will depend on success realizations when the true qualities are not known.

\begin{definition}[Ex-Post Incentive Compatibility]
We say that a mechanism is ex-post incentive compatible if all the bidders are truthful for every success realization irrespective of the bids of other workers, i.e., $\forall i \in \mathcal{N},\ \forall \rho \in \{0,1,-1\}^{n\times T}$
\begin{align*}
&-c_i\mathcal{A}_i(c_i,\hat{c}_{-i},\rho) + \mathcal{P}_i(c_i,\hat{c}_{-i},\rho) \ge -c_i\mathcal{A}_i(\hat{c}_i, \hat{c}_{-i},\rho) + \mathcal{P}_i(\hat{c}_i, \hat{c}_{-i},\rho),\ \forall  \hat{c}_i \in [0,1], \hat{c}_{-i} \in [0,1]^{n-1}\; .
\end{align*}
\end{definition}

\begin{definition}[Ex-Post Individual Rationality]
We say that a mechanism is ex-post individual rational if for every success realization, truth telling does not give negative utility to any player corresponding to any bids of other players,
 \begin{align*}
-\hat{c}_i\mathcal{A}_i(\hat{c}_i,c_{-i},\rho) + \mathcal{P}_i(\hat{c}_i,c_{-i},\rho) \ge 0\; \forall  \hat{c}_i \in [0,1], c_{-i} \in [0,1]^{n-1}, \rho \in \{0,1,-1\}^{n\times T}\; .
\end{align*}
\end{definition}

\begin{definition}[Ex-Post Monotone Allocation Rule]
If the allocation rule is monotone with respect to every success realization then we say that it is ex-post monotone.
\end{definition}

\begin{align}
\label{eq:ex-post_monotone}
 \mathcal{A}_i^t(\hat{c}_i, c_{-i};\rho) \le \mathcal{A}_i^t(c_i,c_{-i};\rho),\;\
 \forall i \in \mathcal{N},\ \forall t \in \{1,2,\dots,T\},\ \forall \hat{c}_i \ge c_i,\ \forall \rho\; .
\end{align}

Based on the above preliminaries for truthful implementation of a MAB algorithm, we now present the strategic version of \CCB-NS algorithm which we call the \CCB-S algorithm.

\subsection{The \CCB-S Algorithm}
\noindent As we have seen in the previous section, we need monotonicity of an allocation rule for incentive compatibility. In CCB-NS, if the selected set $S^t$ in Step \ref{check_constraint} does not satisfy the constraint with $\hat{q}^-$, in Step \ref{additional_set}, we add agents to satisfy the constraint. This step does not consider strategic costs and thus leads to a violation of monotonicity. Hence, we design a new algorithm which achieves the necessary monotonicity.

In order to ensure truthfulness, we modify the CCB-NS algorithm to select all the workers if the constraint is not satisfied with respect to the lower confidence bound (Step \ref{additional_set}). We present CCB-S in Algorithm \ref{alg:ccb_s}. We then show that the allocation rule given by the algorithm is ex-post monotone. Thus, we can apply results from \cite{BABAIOFF10} to achieve an ex-post incentive compatible and ex-post individual rational mechanism. Before going to the formal analysis of CCB-S, we first formally present an important result in \cite{BABAIOFF10}, which is relevant in our setting:

\begin{thm}\cite{BABAIOFF10}\\
\label{thm:babaioff}
 Let $\mathcal{A}$ be a stochastically monotone
(resp., ex-post monotone) MAB allocation rule. There exists a transformation such that the mechanism $\mathcal{M}$ obtained by applying the transformation to the allocation rule $\mathcal{A}$ satisfies the following properties:
(a) $\mathcal{M}$ is stochastically truthful (resp., ex-post truthful), and ex-post individually rational.
(b) For each success realization, the difference in expected welfare between $\mathcal{A}$ and $\mathcal{M}$ is at most $\gamma n$ where $0 < \gamma < 1$ is the parameter provided to the transformation.
\end{thm}

Let $\mathcal{A}^{CCB-S}$ be the allocation induced by the CCB-S. Using Theorem 5.1, we transform it to  $\tilde{\mathcal{A}}^{CCB-S}$ and payment to be $\mathcal{P}^{CCB-S}$. With this we propose a new mechanism \emph{CCB-S}, where $\mathcal{M}^{CCB-S} = \{\tilde{\mathcal{A}}^{CCB-S},\mathcal{P}^{CCB-S}\}$ for the requester to select a subset of 
strategic workers with unknown qualities. We now prove that CCB-S has desirable game theoretic properties.

\begin{algorithm}[h!]{}
\DontPrintSemicolon
\caption{\CCB-S Algorithm}\label{alg:ccb_s}
\KwIn {Set of workers $\mathcal{N}$, number of tasks $T$, parameter $\alpha$, confidence level $\mu$}

\KwOut {Labeler selection set $S^t$, Label $\hat{y}^t$ for all tasks $t \in \{1,2,\ldots,T\}$}
$\forall i \in \mathcal{N}$, $\hat{q}_i^+ = 1$, $\hat{q}_i^- = 0.5$, $k_{i}(1) = 0$ // Initialize UCB and LCB on qualities\;
$S^1 = \mathcal{N}$ // Select all workers initially \label{initialize} \;

Observe $\tilde{y}(S^1)$ and $\hat{y}^1 = \text{AGGREGATE}(\tilde{y}(S^1))$ (Definition \ref{def:aggregate})\; 

Observe true label $y^1$ \;

$\forall i \in \mathcal{N}$, $n_{i}(1) = 1$, $k_{i}(1) = 1$ if $\tilde{y}_i^1 = y^1$ and $\hat{q}_i = k_{i}(1)/n_{i}(1)$ \; 
$t = 2$ \;
$S^t = \displaystyle\argmin_{S \subseteq \mathcal{N}} \displaystyle\sum_{i \in S} c_i\ \text{s.t.}\ f_S(\hat{q}^+) < \alpha$ \;
\While{$f_{S^t}(\hat{q}^-) > \alpha$} 
{ \label{check_constraint}  // Explore (not the optimal set, select all the workers)\;
    $S^t = \mathcal{N}$  \label{additional_set} \;
    Observe judgments of selected labelers $\tilde{y}(S^t)$ \;
    $\hat{y}^t = \text{AGGREGATE}(\tilde{y}(S^t))$\;
    Observe true label $y^t$ \; 
    \For{$i \in S^t$}
	{$n_{i}(t) = n_{i}(t-1) + 1$\; 
    \If{$\tilde{y}_i^t = y^t$}{$k_{i}(t) = k_{i}(t-1) + 1$}
    
    $\hat{q}_i = k_{i}(t)/n_{i}(t)$, $\hat{q}_i^+ = \hat{q}_i + 
    \sqrt{\frac{1}{2n_{i}(t)} \ln(\frac{2n}{\mu})}$, $\hat{q}_i^- = \hat{q}_i 
    - \sqrt{\frac{1}{2n_{i}(t)} \ln(\frac{2n}{\mu})}$ \label{update} \;}
    $t = t+1$\;
    $S^t = \displaystyle\argmin_{S \subseteq \mathcal{N}} \displaystyle\sum_{i \in S} c_i\ \text{s.t.}\ f_S(\hat{q}^+) < \alpha$ \label{selected_set} \;}
	$t^* = t$ \label{start_exploitation_strategic} \;
        $S^{t^*} = S^t$ \label{return_opt_set}\;

\For{$t = t^* + 1$ to $T$\label{start_exploitation}} {
// Exploit (optimal set with high probability)\;
	$S^t = S^{t^*}$ \;
	Observe judgements of selected labelers $\tilde{y}(S^t)$ \;
	$\hat{y}^t = \text{AGGREGATE}(\tilde{y}(S^t))$ \label{exploitation_step} 
}
\end{algorithm}

\subsection{Analysis of $\mathcal{M}^{CCB-S}$}
  \noindent As we are using an exploration-seperated allocation rule,  
one natural way to ensure truthfulness is to apply the classical VCG mechanism. We cannot apply the VCG payment scheme in this algorithm as computing VCG payments requires the computation of an allocation rule in the absence of worker $i$ which cannot be determined by the algorithm since learning stops after computing the optimal set.

In order to design an ex-post incentive compatible and ex-post individual rational mechanism, it is enough to design an ex-post monotone allocation rule. Thus, we will show that our algorithm achieves ex-post monotonicity and hence we can achieve an ex-post truthful and ex-post individual rational mechanism (Theorem \ref{thm:babaioff}).

\begin{thm}
\label{thm:ex-post_monotone}
 The allocation rule given  by the \CCB-S algorithm ($\mathcal{A}^{CCB-S}$) is ex-post monotone.
\end{thm}
\textbf{Proof:} For notation brevity, let us denote $\mathcal{A}^{CCB-S}$ by $\mathcal{A}$. In order to prove monotonicity, we need to prove the following:
\begin{align*}
 \mathcal{A}_i^t(\hat{c}_i, c_{-i};\rho) \le \mathcal{A}_i^t(c_i,c_{-i};\rho),\;\
 \forall i \in \mathcal{N},\ \forall t \in \{1,2,\dots,T\},\ \forall \hat{c}_i \ge c_i,\ \forall \rho\; .
\end{align*}
For a fixed success realization $\rho$, let us denote $\mathcal{A}_i^t(\hat{c}_i, c_{-i};\rho)$ by $\mathcal{A}_i^t(\hat{c}_i, c_{-i})$ for notation brevity.
Since task $t = 1$ is given to all the workers irrespective of their bids, we have $\mathcal{A}_j^1(\hat{c}_i, c_{-i}) = \mathcal{A}_j^1(c_i,c_{-i}) = 1\ \forall j \in \mathcal{N}$. Let $t$ be the largest time step such that, $\forall j$,  $\mathcal{A}_j^{t-1}(\hat{c}_i, c_{-i}) = \mathcal{A}_j^{t-1}(c_i,c_{-i}) = t-1$ (Exploration round with $\hat{c}_i$ and $c_i$). And $\exists i$ such that,
\begin{align*}
&\mathcal{A}_i^t(\hat{c}_i, c_{-i}) \ne \mathcal{A}_i^t(c_i, c_{-i})\; .
\end{align*}
Since other costs and quality estimates are the same, this can happen only when in one case worker $i$ is selected, while in the other case worker $i$ is not selected. Let the two sets of workers selected with $c_i$ and $\hat{c}_i$ be $S(c_i)$ and $S(\hat{c}_i)$ respectively. Since the optimization problem involves cost minimization and quality updates are the same, we have,
\begin{align*}
&\mathcal{A}_i^t(\hat{c}_i, c_{-i}) = t-1\ \text{which implies}\ i \notin S(\hat{c}_i), \\
&\mathcal{A}_i^t(c_i,c_{-i}) = t\ \text{which implies}\ i \in S(c_i)\; .\\
\end{align*}
Since $i \notin S(\hat{c}_i)$, the selected set $S(\hat{c}_i)$ satisfies the lower confidence bound too (exploitation round with bid $\hat{c}_i$) and thus for the rest of the tasks, only $S(\hat{c}_i)$ is selected and thus we have, $\mathcal{A}_i^t(\hat{c}_i, c_{-i}) \le \mathcal{A}_i^t(c_i,c_{-i})$.
\proofends

We denote the mechanism $\mathcal{M}^{CCB-S}=(\mathcal{A}^{CCB-S}, \mathcal{P}^{CCB-S})$. As given by Theorem \ref{thm:babaioff}, $\mathcal{P}^{CCB-S}$ can be derived by applying the transformation given in \cite{BABAIOFF10}. Thus, we obtain the following corollary:
\begin{corollary}
\label{cor:ic_mechanism}
The \CCB-S algorithm produces an ex-post incentive compatible and ex-post individual rational mechanism.
\end{corollary}

One can apply the transformation presented in \cite{BABAIOFF10} which takes any ex-post monotone allocation rule as input and outputs a randomized mechanism which is ex-post incentive compatible and ex-post individually rational.

\subsubsection*{Regret Analysis}

\noindent The proposed algorithm is adaptive exploration separated and the number of exploration steps is determined based on how learning progresses. We will show that $t^*$ returned by Step \ref{start_exploitation_strategic} of the algorithm is bounded by $\frac{2}{(h^{-1}(\Delta))^2}\ln(\frac{2n}{\mu})$. In Lemma \ref{lemma1_non_unif}, we prove that after $l = \frac{2}{(h^{-1}(\Delta))^2}\ln(\frac{2n}{\mu})$ steps, there is no set $S$ which satisfies the constraint with respect to upper confidence bound and its cost is less then the optimal cost. Moreover, after $l = \frac{2}{(h^{-1}(\Delta))^2}\ln(\frac{2n}{\mu})$ steps, we have $f_{S^*}(\hat{q}^-) < \alpha$ with probability $1-\mu$.
\begin{lemma}
\label{lemma1_non_unif}
 After $l = \frac{2}{(h^{-1}(\Delta))^2}\ln(\frac{2n}{\mu})$
number of uniform exploration rounds,
\begin{enumerate}
\item for all sets $S\neq S^*$, $f_S(q) > \alpha \implies f_S(\hat{q}^+) > \alpha$ with probability $1-\mu$\\
\item $f_{S^*}(\hat{q}^-) < \alpha$ with probability $1-\mu$.
\end{enumerate}
\end{lemma}
The proof follows from Lemma \ref{lemma1_non_strat} as after $l$ uniform exploration rounds, we have $n_i(t) \ge l,\ \forall i \in \mathcal{N}$

As a result of Lemmas \ref{opt_set} and \ref{lemma1_non_unif},  we have the following theorem which gives us the bound on the number of non-optimal rounds:

\begin{thm}
\label{thm:bad_rounds}
The number of non-optimal rounds of the \CCB-S algorithm is bounded by $\frac{2}{(h^{-1}(\Delta))^2}\ln(\frac{2n}{\mu})$ with probability $1-\mu$.
\end{thm}
\textbf{Proof:}
\begin{itemize}
\item From Lemma \ref{opt_set}, the CCB-S exploitation rounds are optimal rounds.
\item From Lemma \ref{lemma1_non_unif}, the number of exploration rounds is bounded by $\frac{2}{(h^{-1}(\Delta))^2}\ln(\frac{2n}{\mu})$ with probability $1-\mu$.
\end{itemize}
Hence the theorem follows.
\proofends

\noindent{\em Remark (1):} The algorithm \CCB-S turns out to be an exploration separated algorithm where the number of exploration steps is adaptive unlike the algorithms presented in \cite{DEVANUR09,GATTI12} and bounds on the number of exploration steps  depend on the parameters $\Delta$ and $\mu$.

\noindent{\em Remark (2):} When the value of $\Delta$ is very small compared to $T$ i.e. $(h^{-1}(\Delta))^2 < \frac{1}{T}\ln\left(\frac{2n}{\mu}\right)$, then the algorithm might not converge before $T$ time steps. In a classical MAB algorithm, for example UCB1, there is an inverse dependence on $\Delta_*$ (difference between sub-optimal arm and optimal arm). If $\Delta_*$ is low, then UCB1 suffers a large regret. For practical situations, where $\Delta$ is very low, the requester could provide a range for target accuracy to circumvent high regret. More details are given in Section \ref{sec:simulations}. 

We have the following Corollary that follows from Theorem \ref{thm:bad_rounds}:

\begin{corollary}
 The total expected regret is bounded by
$$\left(1-\frac{1}{T}\right)\frac{2}{(h^{-1}(\Delta))^2}\ln(2nT)C(\mathcal{N}) + L,$$
where $L$ is the loss incurred by the requester if the constraint is not satisfied.
\end{corollary}

\section{Practical Aspects and Experimental Results}
Up until now, our focus was on a combinatorial framework that solves a general optimization problem. A naive implementation of the CCB-NS algorithm and the CCB-S algorithm may lead to two problems: 1) computational complexity of the underlying optimization problem 2) high cost of exploration for the CCB-S algorithm. 

In practice, often the underlying optimization problems are well studied combinatorial problems. Due to this, often we may still be able to use the AAB framework to address the complexity concerns through efficient approximation algorithms that satisfy monotonicity that we define later.  In this section, we consider the majority rule as the aggregation rule and solve Example \ref{ex:min_knapsack} by formulating it as a minimum knapsack problem. The minimum knapsack problem is NP-hard, however, there exists polynomial time greedy approximate algorithm that yields a factor of $2$ approximation for this problem.

To ensure truthfulness, CCB-S selects all the workers in the exploration steps and this might result in very high cost when $n$ is large. In general, it is difficult to eliminate the low quality or high cost workers due to the combinatorial nature of the problem. However, if there exists a structure to the optimization problem, it is often possible to eliminate the workers. 
In the approximate solution of the minimum knapsack problem, we show that it is possible to early identify and eliminate the workers of high cost and low quality in CCB-S algorithm. This elimination avoids high cost of exploration.

\label{sec:practical_aspects}

\subsection{Working with Approximate Solutions}
The key to incorporate approximate algorithm in AAB framework is to show its monotonicity in cost. The CCB-S algorithm that uses the solution returned by the monotone approximate algorithm gives a monotone allocation rule which is essential for incentive compatibility.

\begin{definition}[Monotone Algorithm]
An algorithm is said to be monotone if the allocation $\mathcal{A}$ returned by the algorithm is monotone in cost i.e. if two input instances are
$(c,q)$ and $(c^+,q)$ such that $c_i < c_i^+$, for some $i$ and $c_j = c_j^+\ \forall j \ne i$, then
$\mathcal{A}_i(c^+,q) = 1 \Rightarrow \mathcal{A}_i(c,q) = 1$. 
\end{definition}

\begin{definition}[$(\beta, \gamma)$ Approximate Algorithm]
An algorithm is said to be a $(\beta, \gamma)$ approximate algorithm, if for some $\beta \ge 1$ and $\gamma \le 1$, the solution set $S$ returned by the algorithm is such that $\mathbb{P}[C(S) \le \beta C(S^*)] \ge \gamma$. Here, $S^*$ is the solution returned by optimal algorithm. 
\end{definition}

\begin{proposition}
If there exists a $(\beta,\gamma)$ approximation algorithm that is monotone, then, incorporating that $(\beta, \gamma)$ approximation scheme in the CCB-S algorithm will result in an ex-post monotone allocation rule. 
\end{proposition}
This is easy to see from Theorem \ref{thm:ex-post_monotone}. Note that, all the workers are selected in the exploration rounds, and in exploitation rounds, if a worker $i$ is selected with a certain cost, he will also be selected with a lower cost due to the monotonicity property of the approximate algorithm.

\noindent{\em Remark:} The regret notion in the approximation setting is similar to that in the exact version as the underlying optimization problem with known quality is also solved using the approximate algorithm.

We now present an example of the minimum knapsack optimization problem and a greedy solution of the problem. In the greedy solution, it is possible to eliminate the workers without violating monotonicity condition, thus, avoiding high cost of exploration. We present the elimination strategy for this example. 




\subsection{An Illustrative Example with Low Regret}

From Example \ref{ex:min_knapsack},  if all the workers have  qualities of at least $\frac{2}{3}$, i.e. $q_i > 2/3$ and $\epsilon = 1/6$, then the optimization problem of minimizing cost and satisfying the accuracy constraint of $\alpha$ can be formulated as follows:
\begin{flalign}
&\min_{S \in \mathcal{N}} C(S) \nonumber\\
&\text{s.t.} \sum_{i\in S} (2q_i - 1) \ge 6\ln\left(\frac{1}{\alpha}\right) \nonumber
\end{flalign}

This turns out to be the minimum knapsack problem when $c_i \ge 0$ and $2q_i - 1 \ge 0\ \forall i$. Denote $a_i = 2q_i - 1$ and $M = 6\ln\left(\frac{1}{\alpha}\right)$, we have the following optimization problem:

\begin{flalign}
&\min_{S \in \mathcal{N}} C(S) \label{eq:example}\\
&\text{s.t.} \sum_{i\in S} a_i \ge M \nonumber
\end{flalign}

\subsubsection{Greedy algorithm ($GA$)}
The minimum knapsack problem has a greedy deterministic algorithm which gives a $(2,1)-$approximate solution \cite{CSIRIK90}. The algorithm denoted by $GA$ is as follows:
\begin{itemize}
 \item Arrange the workers in ascending order of their $c_i/a_i$ ratio. Without loss of generality, let us assume that the workers are indexed such that $\frac{c_1}{a_1} \le \frac{c_2}{a_2}\le \ldots \le \frac{c_n}{a_n}$.
 \item Let $k_1$ be the index of the worker such that $\sum_i^{k_1} a_i < M$ but $\sum_i^{k_1} a_i + a_{k_1 + 1} \ge M$. Let $S_0 = \{1,2,\ldots,k_1\}$.
 \item Let $k_2$ be the index of the worker such that $\sum_{i=1}^{k_1} a_i + a_j \ge M\ \forall k_1+1 \le j \le k_2-1$, but $\sum_{i=1}^{k_1} a_i + a_{k_2} < M$. Let $B_0 = \{k_1+1,k_1+2,\ldots,k_2-1\}$.
 \item Let $k_3$ be the index of the worker such that $\sum_i^{k_1} a_i + \sum_{k_2}^{k_3} a_i < M$ but $\sum_i^{k_1} a_i + \sum_{k_2}^{k_3} a_i + a_{k_3+1} \ge M$. Let $S_1 = \{k_2,k_2+1,\ldots, k_3\}$.
 \item In general let $S_l = \{k_{2l}, k_{2l}+1, \ldots, k_{2l+1}\}$ and $B_l = \{k_{2l+1}+1, k_{2l+1}+2, \ldots k_{2l+2}-1\}$, where $k_0 = 1$ and $k_{2l}$ is such that:
 $\sum_{j=0}^l \sum_{i=k_{2j}}^{k_{2j+1}} a_i < M$ but $\sum_{j=0}^l\sum_{i=k_{2j}}^{k_{2j+1}} a_i + a_m \ge M\ \forall k_{2l+1}+1 \le m \le k_{2l+2}-1$.
 \item Among the sets, $S_l \cup \{j\}\ \text{s.t.}\ j \in B_l$, pick the set which has the minimum cost and output that as the solution. 
\end{itemize}

\begin{lemma} \cite{CSIRIK90}.\\
The greedy algorithm $GA$ gives a solution which is $(2,1)-$approximate to the optimal solution i.e.
 $$C(S^{GA}) \le 2C(S^*)$$
 where $S^{GA}$ and $S^*$ are the the solutions returned by the algorithm $GA$ and optimal algorithm respectively.
\end{lemma}

\begin{lemma}
\label{lemma:monotone}
 The allocation rule given by greedy algorithm is monotone in cost i.e. if worker $i$ gets a task with cost $c_i$, he also gets a task with cost $c_i^-$ when the costs and the qualities of the other workers are fixed and $c_i^- < c_i$.
\end{lemma}
\textbf{Proof:}\\
We will prove this case by case. Let worker $i$ be selected with the cost $c_i$. Call the sets $S_0, S_1, \ldots, S_q$ as small sets and the elements of these sets as small elements since the constraint is not satisfied with these elements. Similarly, the sets $B_0,B_1,\ldots,B_q$ are called as big sets and the elements of these sets are called as big elements. Let the set returned by the algorithm be $S_0 \cup S_1 \cup \ldots \cup S_q \cup \{j\}$ where, $j \in B_q$ with cost $c_i$. Now, consider following cases:
\begin{enumerate}
 \item With cost $c_i$, worker $i$ belongs to set $S_l$ where $l \le q$: Now consider the following cases when the cost of worker $i$ is decreased from $c_i$ to $c_i^-$:
 \begin{enumerate}
  \item Worker $i$ remains in $S_l$ but can appear before some other workers in $S_l$. In this case, nothing will change as no other worker has changed positions. Thus, an optimal solution will still be $S_0 \cup S_1 \cup \ldots \cup S_q \cup \{j\}$ since cost has only reduced and worker $i$ will get selected.
  \item Worker $i$ moves to some $S_m$ with $m \le l$. Since $i$ was already in the small set, $S_m$ will remain small. Moreover, all the other workers from small sets till $S_q$ remains small. Thus, the optimal solution will not change and $i$ will get selected.
  \item Note that the worker $i$ can never become a big element by reducing cost.
  \end{enumerate}
  \item With cost $c_i$, worker $j = i$. Thus, $i$ is a big element with cost $c_i$. Again, consider the following cases when worker $i$ changes his bid to $c_i^-$:
  \begin{enumerate}
  \item Worker $i$ becomes big element such that  $i \in B_m$ with $m \le q$. The optimal set will be $S_0 \cup S_1 \cup \ldots \cup S_m \cup \{i\}$ and hence $i$ remains in the solution.
  \item Worker $i$ becomes small such that $i \in S_m$ with $m \le q$. Since $i$ was big till set $S_q$, some worker $k$ from some small set $S_l$ with $l \le q$ will become big. Then, the optimal set will be $S_0 \cup S_1 \cup \ldots S_l \cup \{k\}$ 
  will become optimal and hence $i$ will be selected.
  \end{enumerate}
\end{enumerate}

\subsubsection{Elimination Strategy with Greedy Algorithm $GA$}
 Let us suppose we have (with suitable relabelling)
\begin{equation}
 \frac{c_1}{\hat{a}^{-}_1}  \leq \frac{c_2}{\hat{a}^{-}_2} \leq \ldots \leq \frac{c_k}{\hat{a}^{-}_k},
 \label{eq:sorted}
\end{equation}
where, $\hat{a}_i^- = 2\hat{q}_i^- - 1$.
Let the set $\{1,\ldots, k\}$ be such that a set $S \subseteq \{1,2,\ldots,k\}$ is selected by $GA$, and thus, they meet the accuracy constraint with their lower confidence bounds. 
Further, let us suppose there exists an agent $r \in \{k+1,\ldots, n\}$ such that $\frac{c_k}{\hat{a}^{-}_k} \leq \frac{c_r}{\hat{a}^{+}_r}$ and $c_r \geq c_i, \forall i \in \{1,\ldots,k\}$, then agent $r$ can be discarded ``safely''. By safely, we mean that with qualities known perfectly, $GA$ algorithm has a candidate solution of cost less than  or equal to any candidate solution containing $r$ with probability $(1-\mu)$.\\
\textbf{Proof:}\\
In the run of $GA$ algorithm with true qualities, the elements $1,\ldots, k$ precedes $r$ due to $\frac{c_k}{\hat{a}^{-}_k} \leq \frac{c_r}{\hat{a}^{+}_r}$ with probability $(1-\mu)$. 

As $\{1,\ldots, k\}$  meets the accuracy constraint with LCB, they meet it with true qualities also (with high probability). Therefore, there exists a $p \in \{1,\ldots, k\}$, which belongs to a big set in the run of $GA$ with true qualities. Any candidate set with $p$ in the run will be of the form $\cup_{i=1}^q S_i \cup \{p\}$. Any candidate set with $r$ will be of the form $\cup_{i=1}^l S_i \cup \{r\}$ with $l\geq q$. Therefore, in the run of $GA$, we can ignore any candidate solutions with $r$ and hence $r$ can be dropped safely. This is because $C(\cup_{i=1}^q S_i \cup \{p\}) \le C(\cup_{j=1}^l S_j \cup \{r\})$ as $\cup_{i=1}^q S_i \subseteq \cup_{j=1}^l S_j$ and $c_p \le c_r$.

\begin{lemma}
Algorithm $GA$ with elimination strategy described above produces monotone allocation rule in terms of cost.
\end{lemma}
\textbf{Proof:}\\
Note that in the exploration phase, if the worker $i$ reduces his cost, he can be eliminated at a later stage only and thus the number of allocations in the exploration phase increases. In exploitation phase, the monotonicity is immediate from Lemma \ref{lemma:monotone}.

We call the algorithm $GA$ with elimination strategy as CCB-SE (constrained confidence bound algorithm with strategic elimination). Due to the above Lemma, CCB-SE algorithm can be transformed into incentive compatible mechanism.
\subsection{Simulation Results} \label{sec:simulations}

In this section, we compare the efficacy of the proposed algorithms via simulations. For simulations, we use the minimum knapsack problem described in the previous section which is solved using the greedy algorithm $GA$. We compare the regret of four algorithms namely, CCB-NS, CCB-S, CCB-SE and a variant of the $\varepsilon_t-$greedy algorithm. The $\varepsilon_t-$greedy algorithm \cite{AUER00} solves the classical multi-armed bandit problem which involves the selection of the single best arm. In the classical version, a random arm is explored with probability $\varepsilon_t$ and the optimal arm (with the highest empirical mean) is selected with probability $1-\varepsilon_t$. We extend the algorithm to the \AAB\ setting by exploring all the workers with probability $\varepsilon_t$ and with probability $(1 - \varepsilon_t)$, we select the minimum cost worker subset which meets the target constraint with the empirically estimated qualities. The parameter $\varepsilon_t = \min\{1, \frac{100}{t}\}$  decreases with time so as to give more weight to exploitation than exploration. Note that, the $\varepsilon_t$ algorithm is not strategyproof. 

\begin{figure}[h!]
\centering
 \includegraphics[width=4.2in]{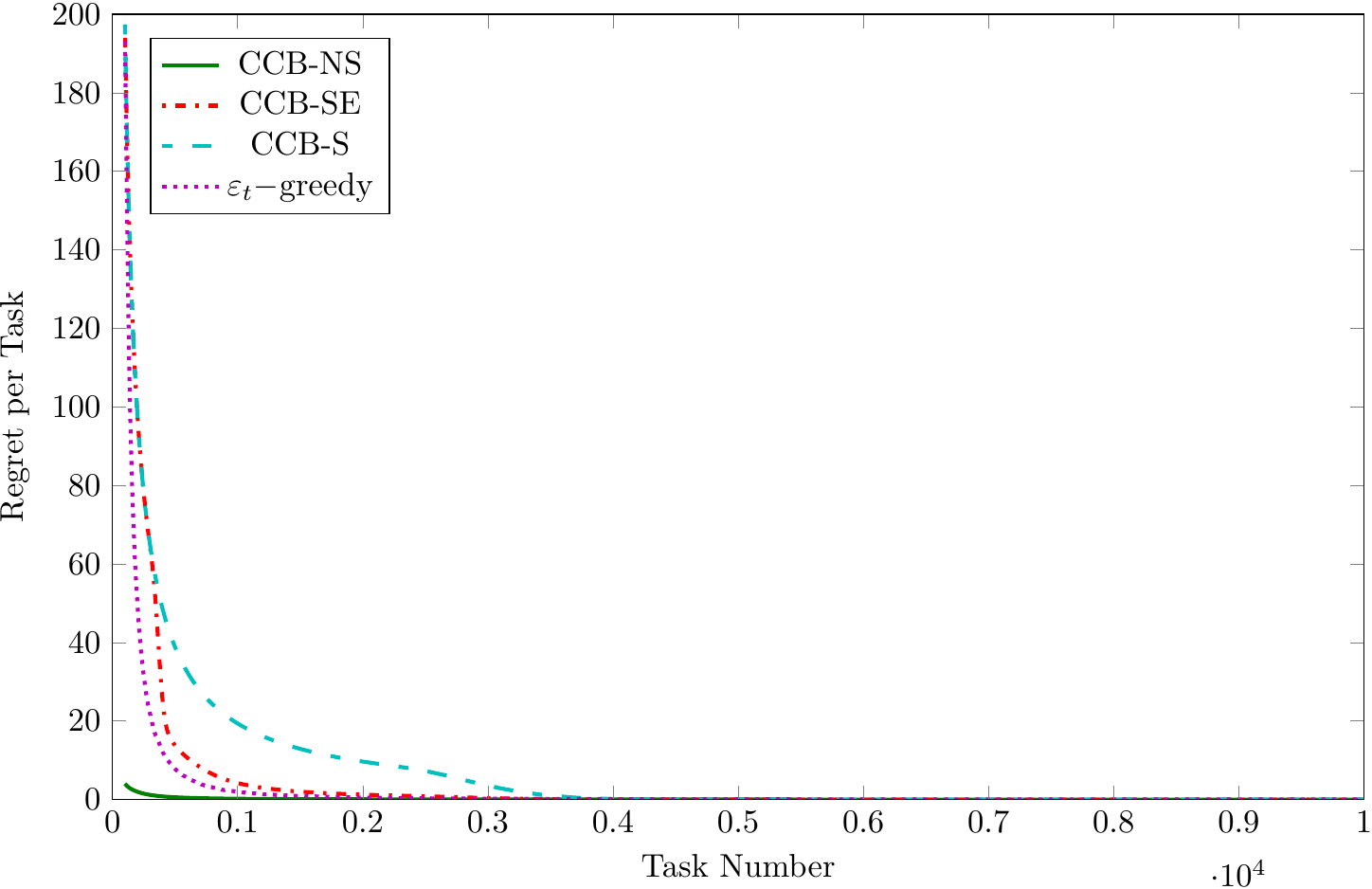}
\caption{Average regret vs task number. Curves show averages over $1000$ samples. Note that CCB-NS and $\varepsilon_t-$greedy algorithms are not ex-post incentive compatible whereas CCB-S and CCB-SE are. }
\label{fig:avg_regret}
\end{figure}

\begin{figure}[h!]
\centering
 \includegraphics[width=4.2in]{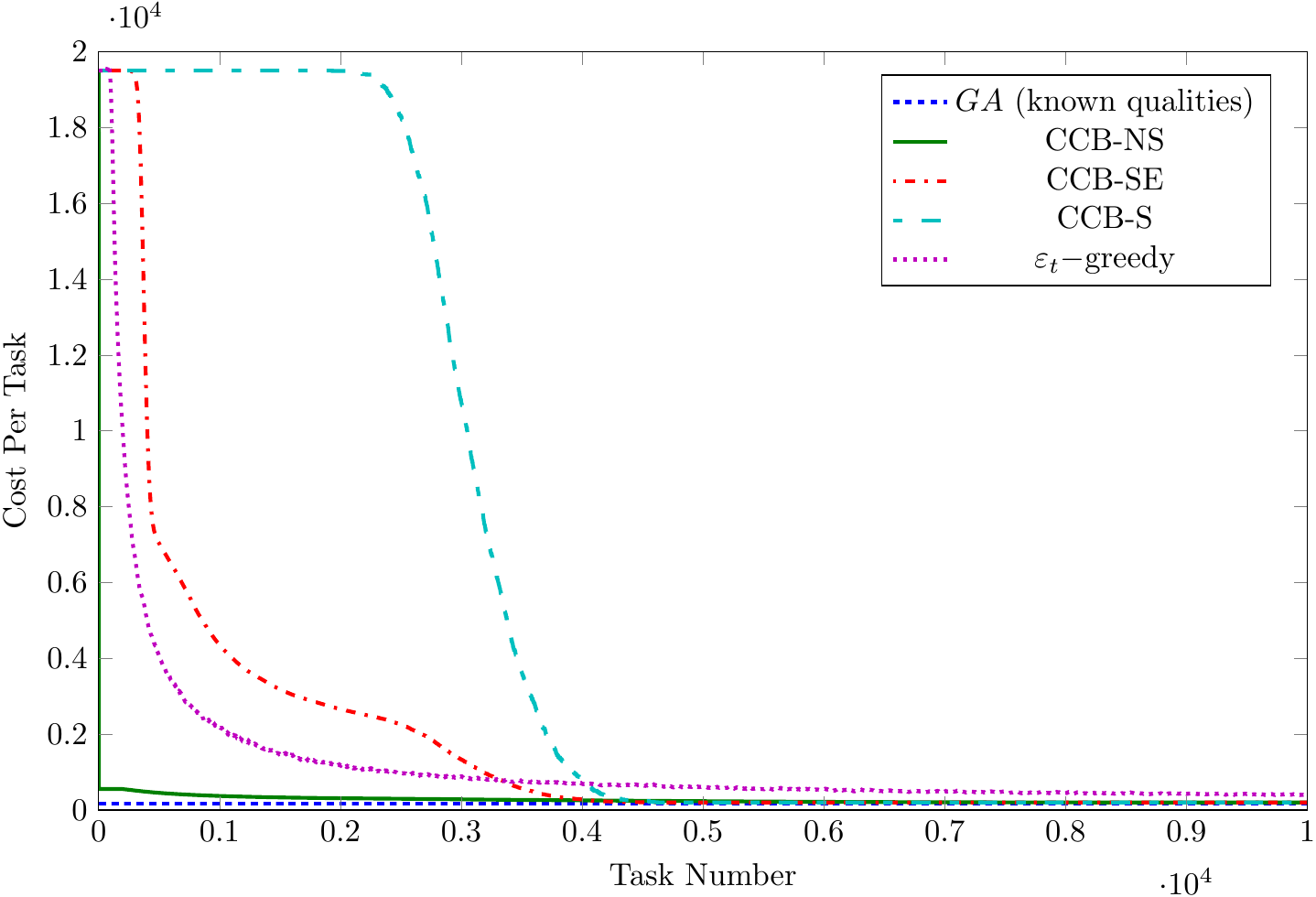}
\caption{Total cost vs task number. Curves show averages over $1000$ samples. Note that CCB-NS and $\varepsilon_t-$greedy algorithms are not ex-post incentive compatible whereas CCB-S and CCB-SE are.}
\label{fig:sw}
\end{figure}

\begin{figure}[h!]
\centering
 \includegraphics[width=4.2in]{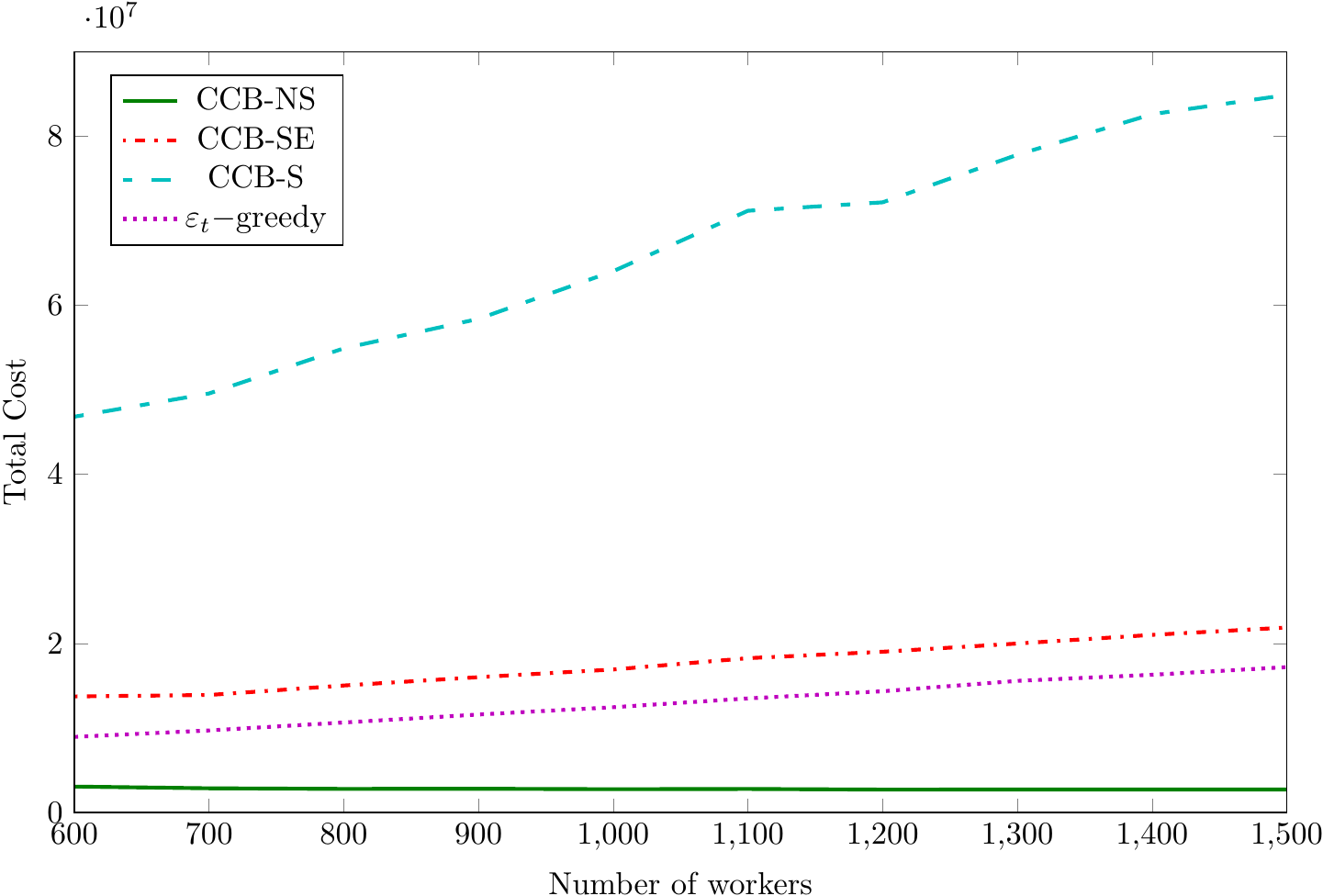}
\caption{Total cost with number of workers}
\label{fig:swvsn}
\end{figure}

In the simulations, we have selected the number of agents to be $1100$. To emphasize the fact that CCB-SE algorithm identifies bad workers early, out of the $1100$ workers, $600$ workers are chosen with cost as $20$ and quality as $2/3$ whereas, the other $500$ workers are chosen with the costs uniformly drawn between $10$ and $20$ and the quality uniformly drawn between $2/3$ and $1$. The required target accuracy is chosen to be $0.9$ with $\alpha = 0.1$. Since the value of $\Delta$ can be arbitrarily low, we adopt the following strategy for the implementation. We solve the optimization problem with UCB for a target accuracy of $0.95$ but check the lower confidence bound with target accuracy $0.9$. This ensures that the constraint is never violated, however, it may result in extra cost of workers for the rest of the rounds. Since the costs of the workers are not adversarially chosen, the expected difference between the optimal set with accuracy $0.9$ and $0.95$ is not large. In general, if the requester gives a target accuracy range of $(1-\alpha, 1-\alpha+\xi)$ such that the upper confidence bound is solved using accuracy $1-\alpha+\xi$ but the lower bound is checked with accuracy $1-\alpha$, then it is possible to control the number of non-optimal rounds and it can be shown that the number of non-optimal rounds is at most $\min\left(\frac{1}{16(h^{-1}(\xi))^2}ln\left(\frac{2n}{\mu}\right), \frac{2}{(h^{-1}(\Delta))^2}ln\left(\frac{2n}{\mu}\right)\right)$. In the $\varepsilon_t-$greedy algorithm, the worker set is chosen such that the target accuracy of $0.9$ is achieved with respect to the estimated qualities. For simulations, we have chosen $T$ to be $10^4$. However, if $T$ is large, one can choose a smaller value of $\xi$. Over $1200$ runs of simulations, we observed that none of the four algorithms violated the stochastic constraint with respect to the true qualities. With $\xi = 0.05$, the comparison of the average regret and the negative social welfare is given in Figures \ref{fig:avg_regret} and \ref{fig:sw} respectively. The regret is compared against the greedy solution returned by $GA$ algorithm with true qualities. We ran $1000$ samples to generate the graphs. We see that the algorithm CCB-NS converges much faster when compared to the $\varepsilon_t-$greedy algorithm. We also see that the cost of CCB-SE algorithm reduces significantly in a few iterations only. We also compare the total cost between CCB-NS algorithm and $\varepsilon_t-$greedy algorithm with change in the number of workers (Figure \ref{fig:swvsn}). The simulations show that the CCB-NS algorithm outperforms the $\varepsilon_t-$greedy algorithm even when there are fewer number of workers. 

\section{Summary and Future Work} \label{sec:future_work}
\noindent Motivated by the need for a mechanism where the worker qualities are not known a priori and their costs are private information, we considered the problem of selecting an optimal subset of the workers so that the outcome obtained from aggregating labels from the selected workers attains a target accuracy level. We proposed a novel framework, Assured Accuracy Bandit (\AAB) in this setting and developed an algorithm, Strategic Constrained Confidence Bound (\CCB-S) for the same, which also leads to an ex-post incentive compatible and ex-post individually rational mechanism. We have provided  bounds on the number of exploration steps that
depends on the target accuracy level and the true qualities.

Often, the optimization problem to be solved for each task inherently has exponential time complexity. In most cases, there exist efficient approximate algorithms for solving the optimization problem. If these algorithms are monotone, then the algorithms can be combined with CCB-S algorithm to provide a truthful, IR mechanism. 

An interesting line of future research could be to improve the convergence rate of CCB-S. The slow convergence of CCB-S can be attributed to the algorithm being exploration separated. If there exists a structure to the algorithm for solving the combinatorial optimization problem, then some strategy for eliminating workers in the strategic setting can be adapted. We have seen a strategy in one example. A generalization of this to all possible optimization problem may require more assumption on the function $f_S(q)$ and forms an interesting future direction. Working with soft constraint formulation of this problem forms another extension for the future.

\bibliographystyle{plain}
\bibliography{crowd.bib}

\begin{thebibliography}{10}

\bibitem{ABRAHAM13}
Ittai Abraham, Omar Alonso, Vasilis Kandylas, and Aleksandrs Slivkins.
\newblock Adaptive crowdsourcing algorithms for the bandit survey problem
  (colt'13).
\newblock In Shai Shalev-Shwartz and Ingo Steinwart, editors, {\em Conference
  On Learning Theory}, volume~30 of {\em JMLR Proceedings}, pages 882--910.
  JMLR.org, 2013.

\bibitem{SHIPRA14}
Shipra Agrawal and Nikhil~R. Devanur.
\newblock Bandits with concave rewards and convex knapsacks.
\newblock In {\em Fifteenth ACM Conference on Economics and Computation
  (EC'14)}, pages 989--1006, 2014.

\bibitem{AUER00}
Peter Auer, Nicol\`{o} Cesa-Bianchi, and Paul Fischer.
\newblock Finite-time analysis of the multiarmed bandit problem.
\newblock {\em Machine Learning}, 47(2-3):235--256, May 2002.

\bibitem{BABAIOFF12}
Moshe Babaioff, Shaddin Dughmi, Robert Kleinberg, and Aleksandrs Slivkins.
\newblock Dynamic pricing with limited supply.
\newblock In {\em Thirteenth ACM Conference on Electronic Commerce (EC'12)},
  pages 74--91. ACM, 2012.

\bibitem{BABAIOFF10}
Moshe Babaioff, Robert~D. Kleinberg, and Aleksandrs Slivkins.
\newblock Truthful mechanisms with implicit payment computation.
\newblock In {\em Eleventh ACM Conference on Electronic Commerce (EC'10)},
  pages 43--52. ACM, 2010.

\bibitem{BABAIOFF09}
Moshe Babaioff, Yogeshwer Sharma, and Aleksandrs Slivkins.
\newblock Characterizing truthful multi-armed bandit mechanisms: extended
  abstract.
\newblock In {\em Tenth ACM Conference on Electronic Commerce (EC'09)}, pages
  79--88. ACM, 2009.

\bibitem{BADANADIYURU12}
Ashwinkumar Badanidiyuru, Robert Kleinberg, and Yaron Singer.
\newblock Learning on a budget: posted price mechanisms for online procurement.
\newblock In {\em Thirteenth ACM Conference on Electronic Commerce (EC'12)},
  pages 128--145. ACM, 2012.

\bibitem{SATYANATH15}
Satyanath Bhat, Shweta Jain, Sujit Gujar, and Yadati Narahari.
\newblock An optimal bidimensional multi-armed bandit auction for multi-unit
  procurement.
\newblock In {\em Proceedings of the 2015 International Conference on
  Autonomous Agents and Multiagent Systems (AAMAS'15)}, pages 1789--1790, 2015.

\bibitem{SATYANATH14}
Satyanath Bhat, Swaprava Nath, Onno Zoeter, Sujit Gujar, Yadati Narahari, and
  Chris Dance.
\newblock A mechanism to optimally balance cost and quality of labeling tasks
  outsourced to strategic agents.
\newblock In {\em Thirtheenth International Conference on Autonomous Agents and
  Multiagent Systems (AAMAS'14)}, pages 917--924, 2014.

\bibitem{BUBECK12}
S{\'e}bastien Bubeck and Nicol{\`o} Cesa-Bianchi.
\newblock Regret analysis of stochastic and nonstochastic multi-armed bandit
  problems.
\newblock {\em Foundations and Trends in Machine Learning}, 5(1):1--122, 2012.

\bibitem{RUGGIERO13}
Ruggiero Cavallo and Shaili Jain.
\newblock Winner-take-all crowdsourcing contests with stochastic production.
\newblock In {\em Proceedings of the first AAAI Conference on Human Computation
  and Crowdsourcing (HCOMP'13)}, 2013.

\bibitem{CHEN14}
Shouyuan Chen, Tian Lin, Irwin King, Michael~R Lyu, and Wei Chen.
\newblock Combinatorial pure exploration of multi-armed bandits.
\newblock In Z.~Ghahramani, M.~Welling, C.~Cortes, N.D. Lawrence, and K.Q.
  Weinberger, editors, {\em Advances in Neural Information Processing Systems
  27}, pages 379--387. Curran Associates, Inc., 2014.

\bibitem{CHEN13}
Wei Chen, Yajun Wang, and Yang Yuan.
\newblock Combinatorial multi-armed bandit: General framework and applications.
\newblock In {\em International Conference on Machine Learning (ICML'13)},
  volume~28, pages 151--159, 2013.

\bibitem{CSIRIK90}
J{\'a}nos Csirik, Johannes Bartholomeus~Gerardus Frenk, Martine Labb{\'e}, and
  Shuzhong Zhang.
\newblock {\em Heuristics for the 0-1 min-knapsack problem}.
\newblock European Institute for Advanced Studies in Management, 1990.

\bibitem{DAWID79}
A.~P. Dawid and A.~M. Skene.
\newblock Maximum likelihood estimation of observer error-rates using the {EM}
  algorithm.
\newblock {\em Journal of the Royal Statistical Society. Series C (Applied
  Statistics)}, 28(1):20--28, 1979.

\bibitem{DEVANUR09}
Nikhil~R. Devanur and Sham~M. Kakade.
\newblock The price of truthfulness for pay-per-click auctions.
\newblock In {\em Tenth ACM Conference on Electronic Commerce (EC'09)}, pages
  99--106, 2009.

\bibitem{DING13}
Wenkui Ding, Tao Qin, Xu-Dong Zhang, and Tie-Yan Liu.
\newblock Multi-armed bandit with budget constraint and variable costs.
\newblock In {\em Twenty Seventh Conference on Artificial Intelligence
  (AAAI'13)}, pages 232--238. AAAI Press, 2013.

\bibitem{DAR06}
Eyal Even-Dar, Shie Mannor, and Yishay Mansour.
\newblock Action elimination and stopping conditions for the multi-armed bandit
  and reinforcement learning problems.
\newblock {\em Journal of Machine Learning Research}, 7:1079--1105, 2006.

\bibitem{FAN15}
Ju~Fan, Guoliang Li, Beng~Chin Ooi, Kian-lee Tan, and Jianhua Feng.
\newblock icrowd: An adaptive crowdsourcing framework.
\newblock In {\em Proceedings of the 2015 ACM SIGMOD International Conference
  on Management of Data}, pages 1015--1030. ACM, 2015.

\bibitem{GARG12}
Dinesh Garg, Sourangshu Bhattacharya, S.~Sundararajan, and Shirish~Krishnaj
  Shevade.
\newblock Mechanism design for cost optimal {PAC} learning in the presence of
  strategic noisy annotators.
\newblock In {\em Twenty Eighth Conference on Uncertainity in Artificial
  Intelligence (UAI'12)}, pages 275--285, 2012.

\bibitem{GATTI12}
Nicola Gatti, Alessandro Lazaric, and Francesco Trov{\`o}.
\newblock A truthful learning mechanism for contextual multi-slot sponsored
  search auctions with externalities.
\newblock In {\em Thirteenth ACM Conference on Electronic Commerce (EC'12)},
  pages 605--622, 2012.

\bibitem{GUJAR15b}
Sujit Gujar and Boi Faltings.
\newblock Auction based mechanisms for dynamic task assignments in expert
  crowdsourcing.
\newblock In {\em Proceedings of the International workshop on Agent Mediated
  E-Commerce and Trading Agent Design and Analysis (AMEC/TADA'15)}, 2015.

\bibitem{GUJAR15a}
Sujit Gujar and Boi Faltings.
\newblock Dynamic task assignments: An online two sided matching approach.
\newblock In {\em Proceedings of the 3rd International workshop on Matching
  Under Preferences (MATCHUP'15)}, 2015.

\bibitem{HO13}
Chien-Ju Ho, Shahin Jabbari, and Jennifer~W. Vaughan.
\newblock Adaptive task assignment for crowdsourced classification.
\newblock In {\em International Conference on Machine Learning (ICML'13)},
  volume~28, pages 534--542, 2013.

\bibitem{SHWETA14}
Shweta Jain, Sujit Gujar, Onno Zoeter, and Y.~Narahari.
\newblock A quality assuring multi-armed bandit crowdsourcing mechanism with
  incentive compatible learning.
\newblock In {\em Thirtheenth International Conference on Autonomous Agents and
  Multiagent Systems (AAMAS'14)}, pages 1609--1610, 2014.

\bibitem{SHIVARAM10}
Shivaram Kalyanakrishnan and Peter Stone.
\newblock Efficient selection of multiple bandit arms: Theory and practice.
\newblock In {\em International Confrence on Machine Learning (ICML'10)}, pages
  511--518, 2010.

\bibitem{KARGER11}
Oh~Sewoong Karger David~R. and Shah Devavrat.
\newblock Budget-optimal crowdsourcing using low-rank matrix approximations.
\newblock In {\em $49^th$ Annual Conference on Communication, Control, and
  Computing (Allerton)}, pages 284--291, 2011.

\bibitem{LAI85}
T.L Lai and Herbert Robbins.
\newblock Asymptotically efficient adaptive allocation rules.
\newblock {\em Advances in Applied Mathematics}, 6(1):4 -- 22, 1985.

\bibitem{LI13}
Hongwei Li, Bin Yu, and Dengyong Zhou.
\newblock Error rate bounds in crowdsourcing models.
\newblock In {\em arXiv preprint arXiv:1307.2674}, 2013.

\bibitem{MYERSON81}
Roger~B. Myerson.
\newblock Optimal auction design.
\newblock {\em Mathematics of Operations Research}, 6(1):pp. 58--73, 1981.

\bibitem{RAYKAR10}
Vikas~C Raykar, Shipeng Yu, Linda~H Zhao, Gerardo~Hermosillo Valadez, Charles
  Florin, Luca Bogoni, and Linda Moy.
\newblock Learning from crowds.
\newblock {\em The Journal of Machine Learning Research (JMLR'10)},
  11:1297--1322, 2010.

\bibitem{SUJIT12}
Akash~Das Sharma, Sujit Gujar, and Y.~Narahari.
\newblock Truthful multi-armed bandit mechanisms for multi-slot sponsored
  search auctions.
\newblock {\em Current Science}, Vol. 103 Issue 9:1064--1077, 2012.

\bibitem{SINGER13}
Yaron Singer and Manas Mittal.
\newblock Pricing mechanisms for crowdsourcing markets.
\newblock In {\em Twenty Second Internation World Wide Web Conference
  (WWW'13)}, pages 1157--1166, 2013.

\bibitem{SINGLA13}
Adish Singla and Andreas Krause.
\newblock Truthful incentives in crowdsourcing tasks using regret minimization
  mechanisms.
\newblock In {\em Twenty Second International World Wide Web Conference
  (WWW'13)}, pages 1167--1178, 2013.

\bibitem{LONG12}
Long Tran-Thanh, Archie~C. Chapman, Alex Rogers, and Nicholas~R. Jennings.
\newblock Knapsack based optimal policies for budget-limited multi-armed
  bandits.
\newblock In {\em Twenty-Sixth Conference on Artificial Intelligence
  (AAAI'12)}, pages 1134--1140, 2012.

\bibitem{VIAPPIANI11}
Paolo Viappiani, Sandra Zilles, HowardJ. Hamilton, and Craig Boutilier.
\newblock Learning complex concepts using crowdsourcing: A {Bayesian Approach}.
\newblock In {\em Algorithmic Decision Theory}, volume 6992, pages 277--291.
  2011.

\bibitem{WITKOWSKI13}
Jens Witkowski, Yoram Bachrach, Peter Key, and David~C. Parkes.
\newblock {Dwelling on the Negative: Incentivizing Effort in Peer Prediction}.
\newblock In {\em Proceedings of the first AAAI Conference on Human Computation
  and Crowdsourcing (HCOMP'13)}, pages 1--8, 2013.

\bibitem{ZHOU14}
Yuan Zhou, Xi~Chen, and Jian Li.
\newblock Optimal pac multiple arm identification with applications to
  crowdsourcing.
\newblock In {\em Proceedings of the 31st International Conference on Machine
  Learning (ICML-14)}, pages 217--225, 2014.

\end{thebibliography}

\end{document}